\documentclass[runningheads]{llncs}
\usepackage{graphicx}

\usepackage{tikz}
\usepackage{comment}
\usepackage{amsmath,amssymb} %
\usepackage{color}

\usepackage[pagebackref,breaklinks,colorlinks]{hyperref}

\usepackage{url}            %
\usepackage{booktabs}       %

\usepackage{changepage}   %
\usepackage{multirow}
\usepackage[percent]{overpic}
\usepackage{enumitem}
\usepackage{varwidth}
\usepackage{listings}
\usepackage{blindtext}
\usepackage{wrapfig}
\usepackage{enumitem}
\usepackage{xcolor,colortbl}
\usepackage{bm}
\usepackage{subcaption}
\usepackage{xspace}
\usepackage{placeins}
\usepackage{stfloats}

\usepackage[export]{adjustbox}

\makeatletter
\renewcommand\paragraph{\@startsection{paragraph}{4}{\parindent}%
  {0pt}
  {-\parindent}
  {\bfseries}}
\makeatother

\usepackage[capitalize]{cleveref}
\crefname{section}{Sec.}{Secs.}
\Crefname{section}{Section}{Sections}
\Crefname{table}{Table}{Tables}
\crefname{table}{Tab.}{Tabs.}

\usepackage{makecell}

\newcommand*\rot{\rotatebox{90}}
\newcommand*\rotslant{\rotatebox[origin=r]{90}}

\newcommand{\sce}[1]{\textsc{e}{#1}}

\definecolor{tablegreen}{rgb}{0.09, 0.45, 0.27}

\definecolor{coloursrgb}{rgb}{0.283,0.131,0.449}
\definecolor{collearnedrgb}{rgb}{0.320,0.771,0.411}
\definecolor{colnonlearnedrgb}{rgb}{0.177,0.438,0.558}
\definecolor{colablrgb}{rgb}{0.119512, 0.607464, 0.540218}

\definecolor{yesgreen}{rgb}{0.07,0.49,0.07}
\definecolor{nored}{rgb}{0.98,0.13,0.12}

\newcommand{\colours}[1]{\textcolor{coloursrgb}{#1}}
\newcommand{\collearned}[1]{\textcolor{collearnedrgb}{#1}}
\newcommand{\colnonlearned}[1]{\textcolor{colnonlearnedrgb}{#1}}
\newcommand{\colabl}[1]{\textcolor{colablrgb}{#1}}

\newcommand{\Eimg}{E_I}
\newcommand{\Gimg}{G_I}
\newcommand{\Dimg}{D_I}

\makeatletter
\DeclareRobustCommand\onedot{\futurelet\@let@token\@onedot}
\def\@onedot{\ifx\@let@token.\else.\null\fi\xspace}

\def\eg{\emph{e.g}\onedot} \def\Eg{\emph{E.g}\onedot}
\def\ie{\emph{i.e}\onedot} 
 
\def\etc{\emph{etc}\onedot}

\def\etal{\emph{et al}\onedot}
\makeatother

\newcommand{\Eres}{E_\textit{res}}
\newcommand{\Gres}{G_\textit{res}}
\newcommand{\Eflow}{E_\textit{flow}}
\newcommand{\Gflow}{G_\textit{flow}}
\newcommand{\Dp}{D_P}

\newcommand{\yimg}{y_I}
\newcommand{\yflow}{y_{t,f}}
\newcommand{\yres}{y_{t,r}}
\newcommand{\DcondI}{\yimg}
\newcommand{\DcondP}{\yres}

\newcommand{\Uflow}{\textit{UFlow}\xspace}

\newcommand{\name}{Ours\xspace}
\newcommand{\namebl}{no-GAN\xspace}
\newcommand{\namecvpr}{SSF\xspace}

\newcommand{\nameIT}{\textit{\name}\xspace}
\newcommand{\nameblIT}{\textit{no-GAN}\xspace}

\newcommand{\loss}[1]{\mathcal{L}_{#1}}

\newcommand{\vsqueeze}{\vspace{-.5em}}

\newcommand{\mysection}[1]{\section{#1} \vspace{-0em}}
\newcommand{\mysubsection}[1]{\subsection{#1} \vspace{-0em}}

\newcommand{\cropinside}[5][]{%
    \node[draw=white!40, very thick, inner sep = 0pt,anchor=north east,outer sep=1pt,#1,yshift=-1.5pt] (#4) at (#5) {%
        \includegraphics[width=0.2\linewidth]{#3}};
    \node[fill=white,anchor=north east,inner sep=1pt,outer sep=1pt, minimum height=10pt,yshift=-0pt] at (#4.north east) {#2};}
    
\newcommand{\cropinsidewest}[5][]{%
    \node[draw=white!40, very thick, inner sep = 0pt,anchor=north west,outer sep=1pt,#1,xshift=1.5pt] (#4) at (#5) {%
        \includegraphics[width=0.2\linewidth]{#3}};
    \node[fill=white,anchor=north east,inner sep=1pt,outer sep=1pt, minimum height=10pt,yshift=-0pt] at (#4.north east) {#2};}

\newcommand{\fullshardlabel}[1]{\phantom{\textsuperscript L}#1\phantom{\textsuperscript L}}

\newcommand{\fullshardraw}[8][]{%
    \node[inner sep = 0pt,outer sep=0pt,anchor=south west] (image) at (0,0) {%
            \includegraphics[width=1\linewidth,#1]{#2}};
    \begin{scope}[x={(image.south east)},y={(image.north west)}]
        \def\temp{#7} \ifx\temp\empty %
        \else \node[fill=white,anchor=north west,inner sep=1pt,outer sep=0pt, minimum height=10pt] at (0,1) {\fontsize{4}{5}{#7}};
        \fi
        \draw[white, very thick, outer sep=2pt] (#5,0) -- (#6,1);
        \node[fill=white,anchor=south east,inner sep=1pt,outer sep=0pt, minimum height=10pt, xshift=-5pt] at (#5,0) {\fullshardlabel{#4}};
        \begin{scope}
            \clip (#5,0) -- (#6,1) -- (1,1) -- (1,0) -- cycle;
            \node[anchor=north east,inner sep=0pt, outer sep=0pt] at (1,1) {\includegraphics[width=1\linewidth,#1]{#3}};
        \end{scope}
        \node[fill=white,anchor=south west,inner sep=1pt,outer sep=0pt, minimum height=10pt, xshift=5pt] at (#5,0) {\fullshardlabel{Original}};
        #8
    \end{scope}}

\begin{document}
\pagestyle{headings}
\mainmatter
\def\ECCVSubNumber{4802}  %

\title{Neural Video Compression using GANs for Detail Synthesis and Propagation}
%

%
%

%
%
\titlerunning{Neural Video Compression using GANs}
\author{Fabian Mentzer\thanks{Equal contributions.} \and
Eirikur Agustsson$^*$ \and
Johannes Ball\'e \and
David Minnen \and
Nick Johnston \and
George Toderici}
\authorrunning{Mentzer and Agustsson et al.}
\institute{Google Research, correspondence to: \email{mentzer@google.com}\vspace{-1em}}
\maketitle

\begin{abstract}
We present the first neural video compression method based on generative adversarial networks (GANs). Our approach significantly outperforms previous neural and non-neural video compression methods in a user study, setting a new state-of-the-art in visual quality for neural methods. We show that the GAN loss is crucial to obtain this high visual quality. Two components make the GAN loss effective: we i) synthesize detail by conditioning the generator on a latent extracted from the warped previous reconstruction to then ii) propagate this detail with high-quality flow. We find that user studies are required to compare methods, i.e., none of our quantitative metrics were able to predict all studies. We present the network design choices in detail, and ablate them with user studies.
\vspace{-1ex}
\keywords{Neural Video Compression, GANs\vspace{-2em}}
\end{abstract}

\vspace{-2ex}
\mysection{Introduction}

\begin{figure}
\tiny
  \begin{tikzpicture}
    \fullshardraw[trim={0 10mm 0 10mm},clip]{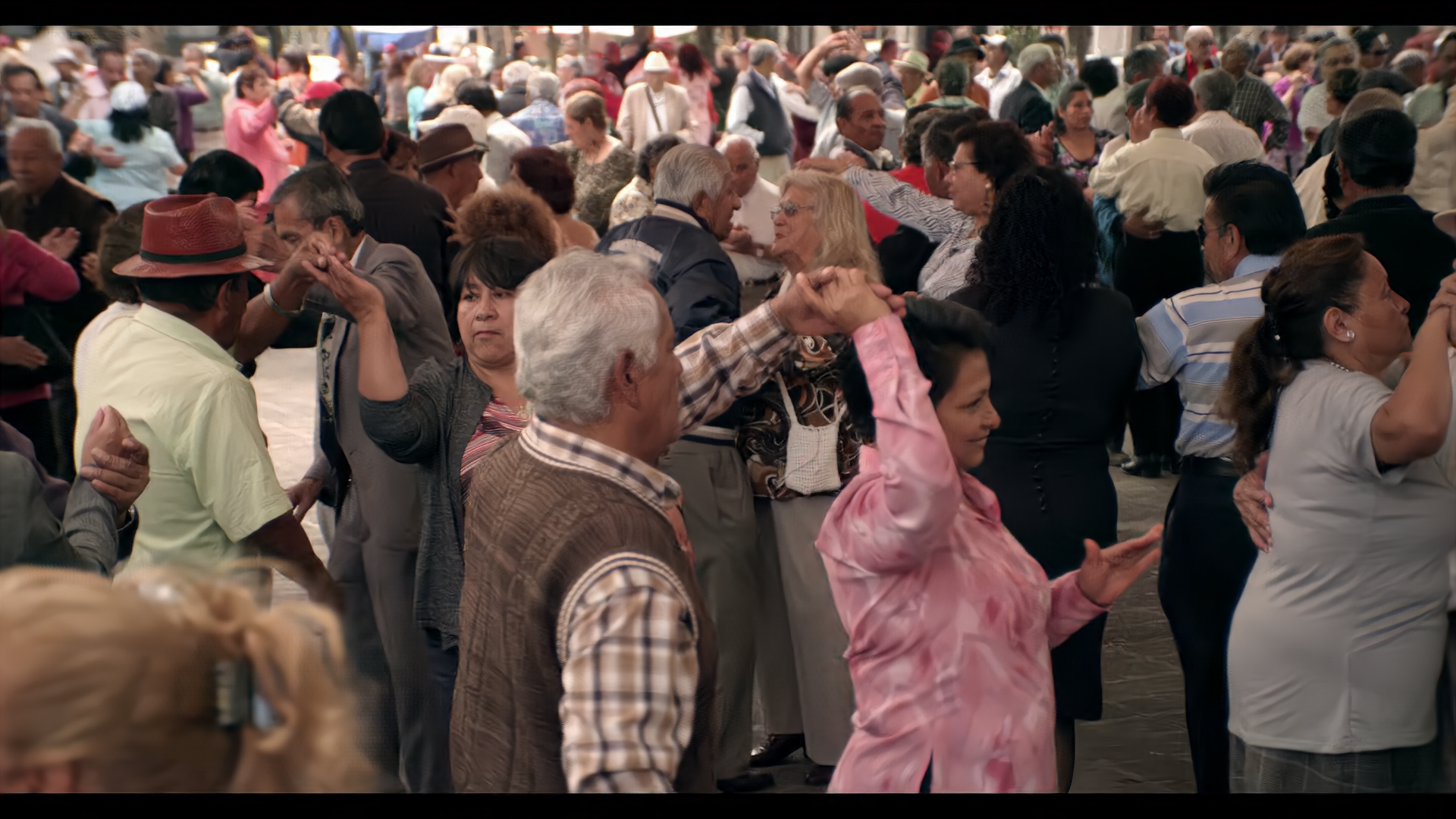}{%
                  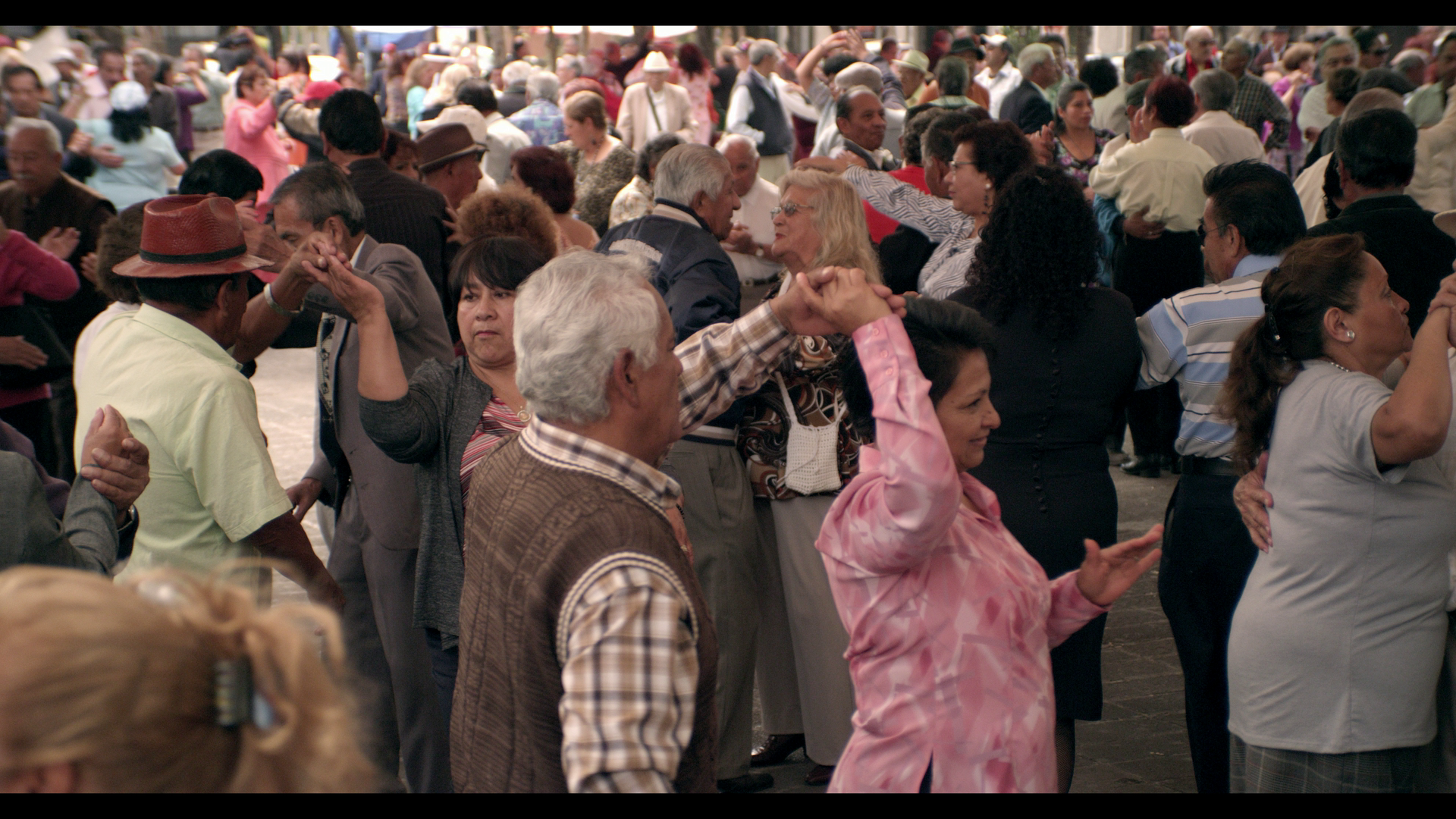}{%
                  Ours}{0.4}{0.65}{}{
        \cropinside[xshift=-1.5pt]{Original}{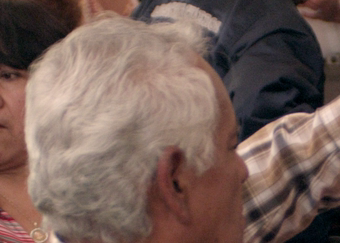}{cropra}{1,1}
        \cropinside{HEVC}{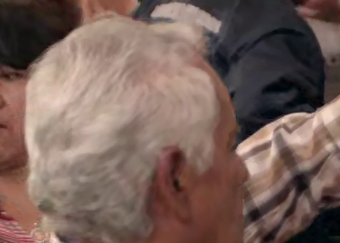}{croprb}{cropra.south east}
        \cropinside{H264}{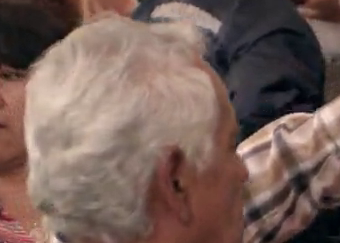}{croprc}{croprb.south east}
        \cropinside[anchor=north west,xshift=1.5pt]{%
          Ours}{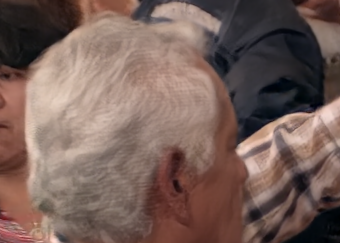}{cropla}{0,1}
        \cropinside{DVC}{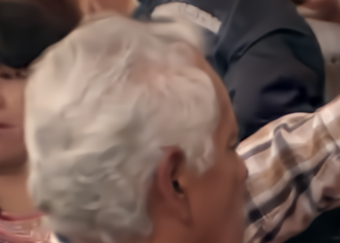}{croplb}{cropla.south east}
        \cropinside{SSF}{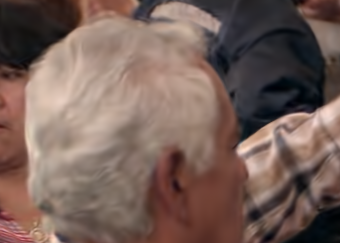}{croplc}{croplb.south east}
        \cropinsidewest{RLVC}{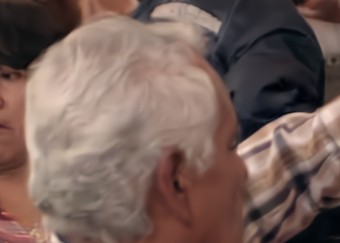}{cropld}{cropla.north east}
        }
  \end{tikzpicture}
\vspace{-4ex}
\caption{\label{fig:frontpic}Comparing our reconstruction to various baselines. 
On the left, we see crops from \emph{neural} methods, where we 
compare to the published MSE-based methods
\collearned{RLVC}~\cite{yang2020learning}, 
\collearned{SSF}~\cite{agustsson2020scale}, and 
\collearned{DVC}~\cite{lu2019dvc}.
On the right we see the original and the non-neural methods, \colnonlearned{H.264}~\cite{AVC} and \colnonlearned{HEVC}~\cite{HEVC}. We see how high frequency texture is faithfully synthesized in our approach, while staying close to the input, where-as MSE-based methods suffer from blurryness. \emph{Best viewed on screen.}
}
\end{figure}

Recently, there has been progress in neural video compression, leading 
to the latest approaches being comparable to or even outperforming the non-learned standard codec HEVC~\cite{HEVC} in terms of PSNR~\cite{agustsson2020scale,yang2020hierarchical,rippel2021elf,li2021deep} or outperforming it in MS-SSIM~\cite{golinski2020feedback,rippel2021elf,li2021deep}.
However, as we navigate the rate-distortion trade-off towards low bitrates, reconstructions become blurry (for neural approaches) or blocky (for non-neural). This was also observed for images, where there has been interest in instead optimizing the rate-distortion-realism trade-off~\cite{blau2019rethinking,tschannen2018deep,theis2021coding,theis2021advantages}. In short, the goal is to add a realism constraint, forcing the decoder to make sure that reconstructions are also looking ``realistic'' (in the sense that they are indistinguishable from real images), while still staying close to the input.
To optimize this constraint, 
previous work~\cite{agustsson2019extreme,mentzer2020high,santurkar2017generative,rippel17a}
added a GAN~\cite{goodfellow2014generative} loss to the rate distortion objective, thereby navigating the triple-tradeoff.

However, targeting realism remains largely unexplored for neural \emph{video} compression.
This is perhaps not surprising, as video compression brings various challenges~\cite{yang2022introduction}, and GAN training is notoriously hard~\cite{goodfellow2014generative}.
To apply rate-distortion-realism theory for video,
we need to be able to synthesize detail whenever new content appears, and then we need to propagate this detail to future frames. 
With this in mind, we carefully design a \emph{generative} neural video compression approach excelling at synthesizing and then preserving detail. 

According to the theory~\cite{blau2018pd,blau2019rethinking}, realism cannot be measured in terms of pair-wise distortions such as PSNR and MS-SSIM. In fact, theory predicts that these metrics must get worse as realism increases.
Following previous work~\cite{agustsson2019extreme,mentzer2020high}, we thus perform extensive user studies to evaluate our approach, where we ask raters to compare methods and chose which ``is closest to the original'' (see Sec.~\ref{sec:userstudy}). 
We find that by trading-off just a little bit in PSNR (${\approx}0.6\text{dB}$, see Sec.~\ref{sec:results}), we can significantly improve in realism, as measured by the study.
This way, our approach manages to synthesize small scale detail while staying close to the original (see Fig.~\ref{fig:frontpic}).
Our main contributions are as follows:

\setlist{nolistsep}\begin{enumerate}[leftmargin=*,noitemsep]
\item{We present the first GAN-based neural compression system and set a new state-of-the-art in subjective visual quality measured with user studies, where we significantly outperform previous neural compression systems (\hspace{1sp}\cite{agustsson2020scale}, \cite{yang2020learning}, \cite{lu2019dvc}), as well as the standard codecs H.264~\cite{AVC} and HEVC~\cite{HEVC}. We show that the GAN loss is crucial for this performance.}
\item{We show that two components are crucial to make the GAN loss effective:
i) We condition the generator on a ``free'' (in terms of bits) latent obtained by feeding the warped previous reconstruction through the image encoder, and show that this is crucial to \emph{synthesize} details. ii) To be able to \emph{propagate} previously synthesized details, we rely on accurate optical flow provided by $\Uflow$~\cite{jonschkowski2020matters}, 
and warping with high-quality resampling kernels.%
}

\end{enumerate}

\mysection{Related Work}
\paragraph{Neural Video Compression}
Wu~\etal~\cite{wu2018video} use frame interpolation for video compression, compressing B-frames by interpolating between other frames. Djelouah~\etal~\cite{djelouah2019neural} also use interpolation, but additionally employ an optical flow predictor for warping frames. This approach of using future frames is commonly referred to as ``B-frame coding'' for ``bidirectional prediction''. Other neural video coding methods rely on only using predictive (P) frames, commonly referred to as the ``low-delay'' setting, since it is more suitable for streaming applications by not relying on future frames.
Lu~\etal~\cite{lu2019dvc} use previously decoded frames and a pretrained optical flow network. 
Habibian~\etal~\cite{habibian2019video} do not explicitly model motion, and instead rely on a 3D autoregressive entropy model to capture spatial and temporal correlations. Liu~\etal~\cite{liu2019neural} build temporal priors via LSTMs, while Liu~\etal~\cite{liu2020conditional} condition entropy models on previous frames. 
Rippel~\etal~\cite{rippel2019learned} support adapting the rate during encoding, and also do not explicitly model motion.
Agustsson~\etal~\cite{agustsson2020scale} propose ``scale-space flow'' to avoid complex residuals by allowing the model to blur as needed via a pyramid of blurred versions of the image.
Yang~\etal~\cite{yang2020hierarchical} generalize various approaches by learning to adapt the residual scale, and conditioning residual entropy models on flow latents.
Li~\etal~\cite{li2021deep} use deep features as context for encoding, decoding and entropy coding.
Golinsky~\etal~\cite{golinski2020feedback} recurrently connect decoders with subsequent unrolling steps, while Yang~\etal~\cite{yang2020learning} also add recurrent entropy models.
Rippel and Anderson~\etal~\cite{rippel2021elf} explore ways to make neural video compression more practical, with models that cover a range of bitrates and a focus on computational efficientcy, improving encode and decode time.

\begin{figure*}[t]
    \centering
    \includegraphics[width=\textwidth]{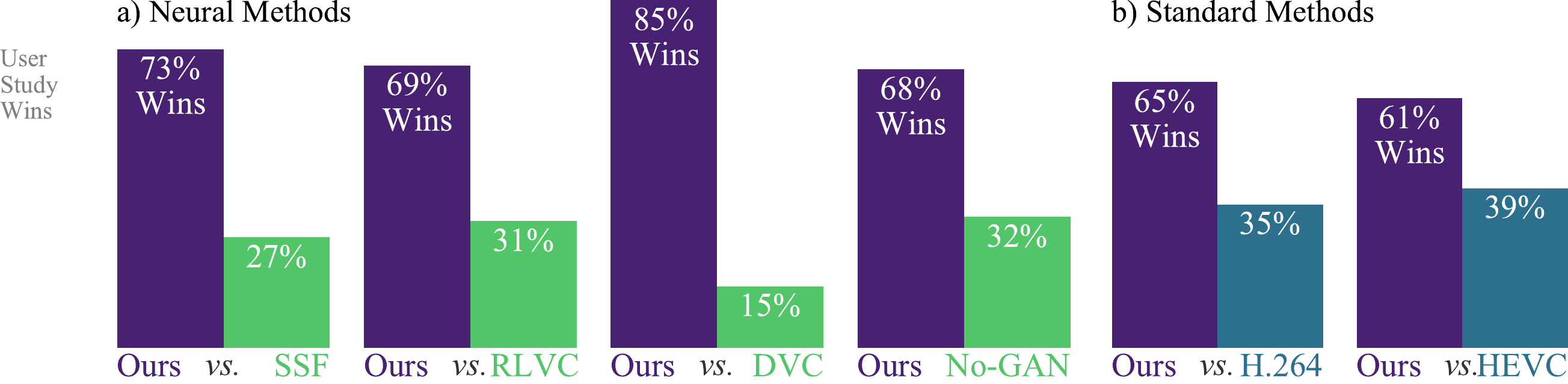}\vspace{-1ex}
    \caption{
    \label{fig:user_study_comparison}
    Comparing 6 different pairs of methods in a \textit{user study}, on MCL-JCV.
    We visualize how often each method is preferred in the user studies.
    We had 1639 ratings in total, with an average of 273 per method pair.
    \emph{a)} Shows neural methods,  \collearned{SSF}~\cite{agustsson2020scale}, 
    \collearned{RLVC}~\cite{yang2020learning}, and
    \collearned{DVC}~\cite{lu2019dvc}, seeing that our method significantly outperforms them in terms of visual quality. 
    We compare \colours{\nameIT} to our 
    \collearned{\nameblIT} baseline, where we see that a GAN clearly helps. 
    \emph{b)} We compare to the standard codecs \colnonlearned{H.264}~\cite{AVC} and \colnonlearned{HEVC}~\cite{HEVC},
    and see that our method is also preferred. 
    \vspace{-3ex}
    }
\end{figure*}

\paragraph{Non-Neural Video Compression}
The combination of transform coding~\cite{Go01} using discrete cosine transforms~\cite{AhNaRa74} with spatial and/or temporal prediction, known as ``Hybrid video coding'', emerged in the 1980s as the technology dominating video compression until the present day.
Non-neural methods such as H.261 through H.265/HEVC~\cite{HEVC}, VP8~\cite{bankoski2011technical}, VP9~\cite{mukherjee2013latest} and AV1~\cite{chen2018overview} have all remained faithful to the hybrid coding principle, with extensive refinements, regarding more flexible pixel formats (\eg, bit depth, chroma subsampling), more flexible temporal and spatial prediction (\eg, I-, P-, B-frames, intra block copy), and many more.  Thanks to the years of research that went into these codecs, they provide strong baselines for neural approaches. 

\definecolor{Gray}{gray}{0.85}
\addtolength{\tabcolsep}{-0.1ex}    
\begin{figure}[t]
    \centering
        {\footnotesize
        \begin{tabular}{>{\columncolor{Gray}}cc@{}}
        & \multirow{6}{*}{
            \includegraphics[width=0.8\linewidth]{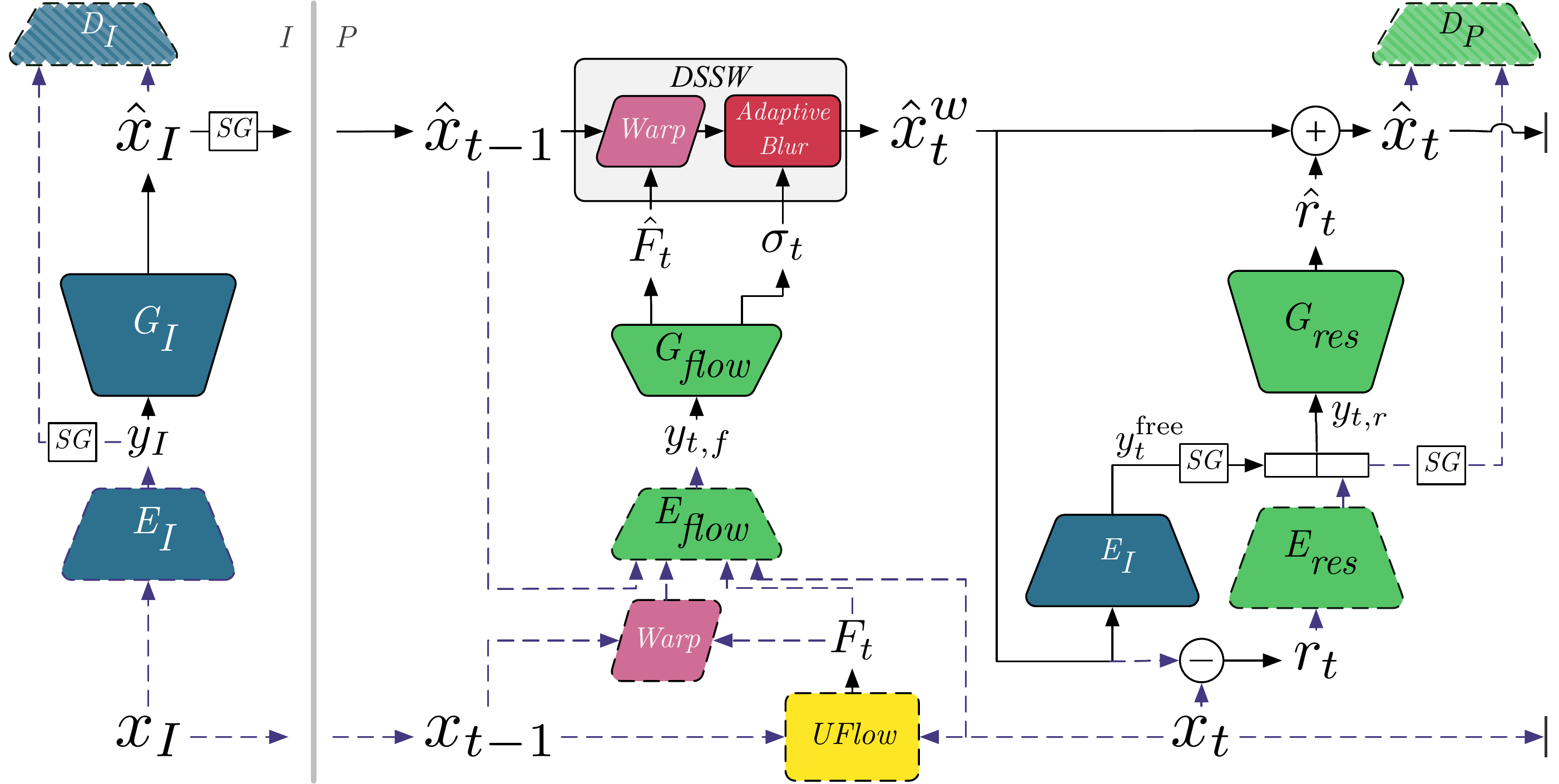}} \\[-2ex]
        \includegraphics[width=0.19\linewidth]{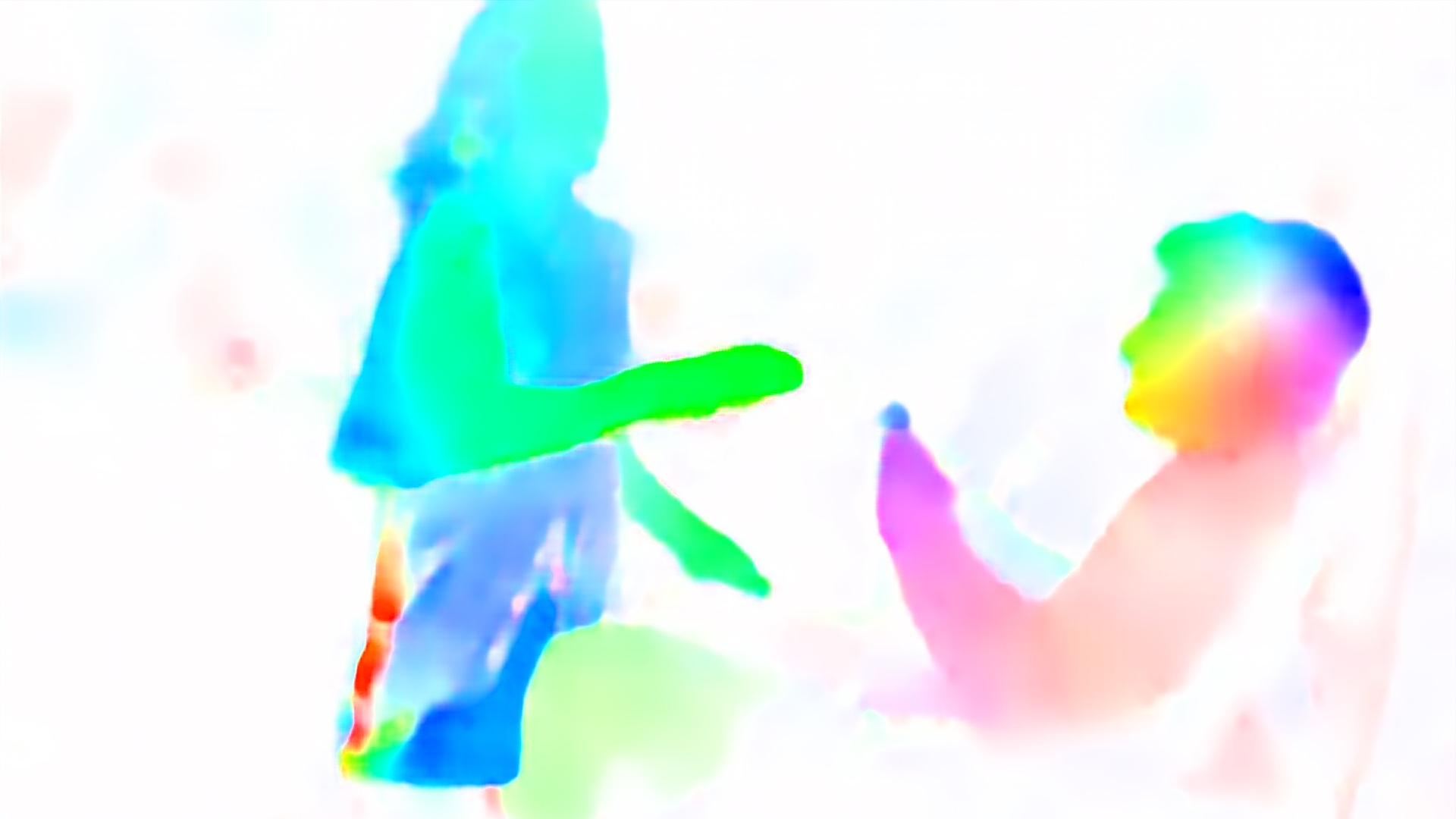}  \\[-0.75ex]
        $F_t$ \\
        \includegraphics[width=0.19\linewidth]{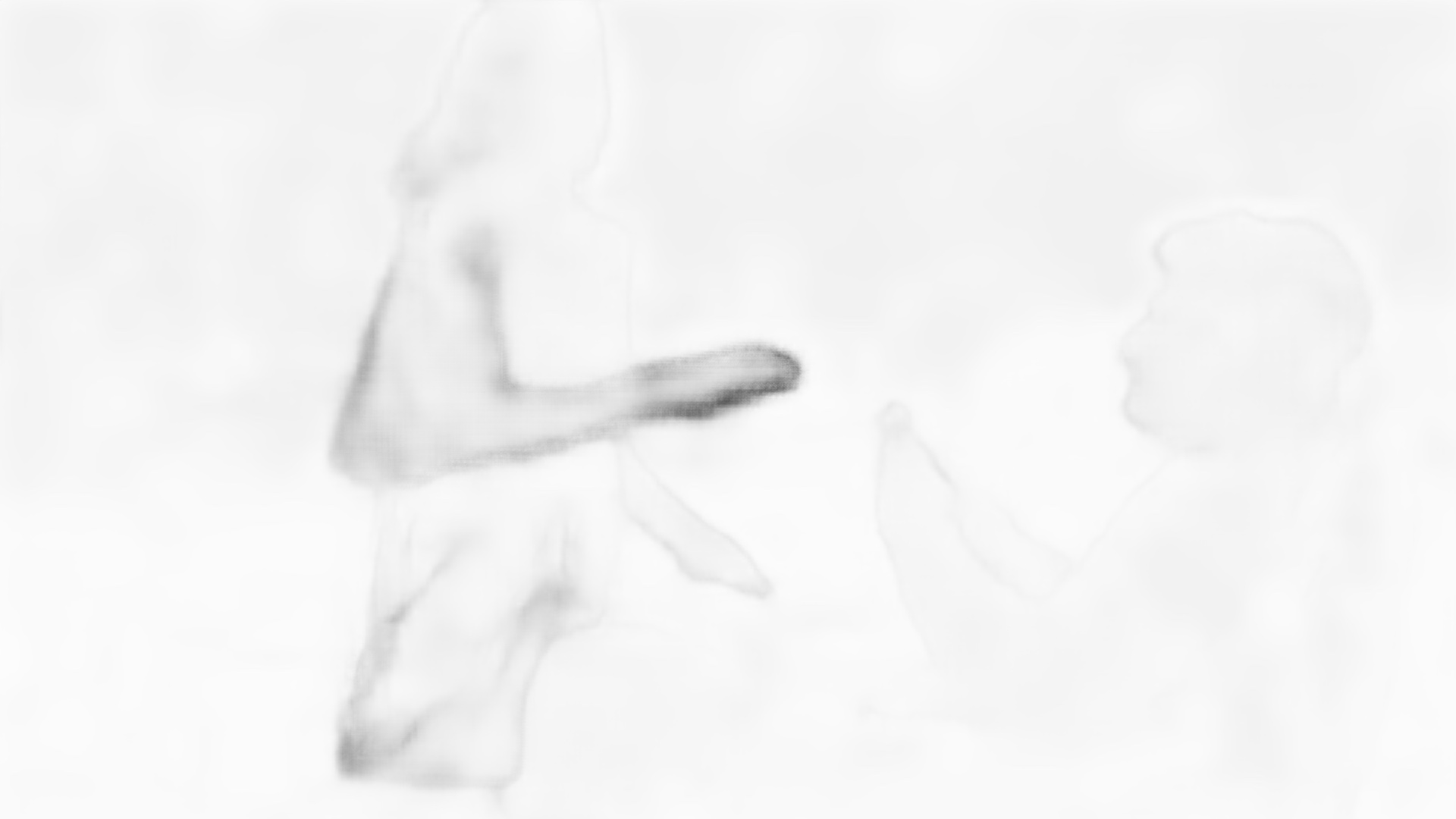}  \\[-0.75ex]
        $\sigma_t$ \\
        \includegraphics[width=0.19\linewidth]{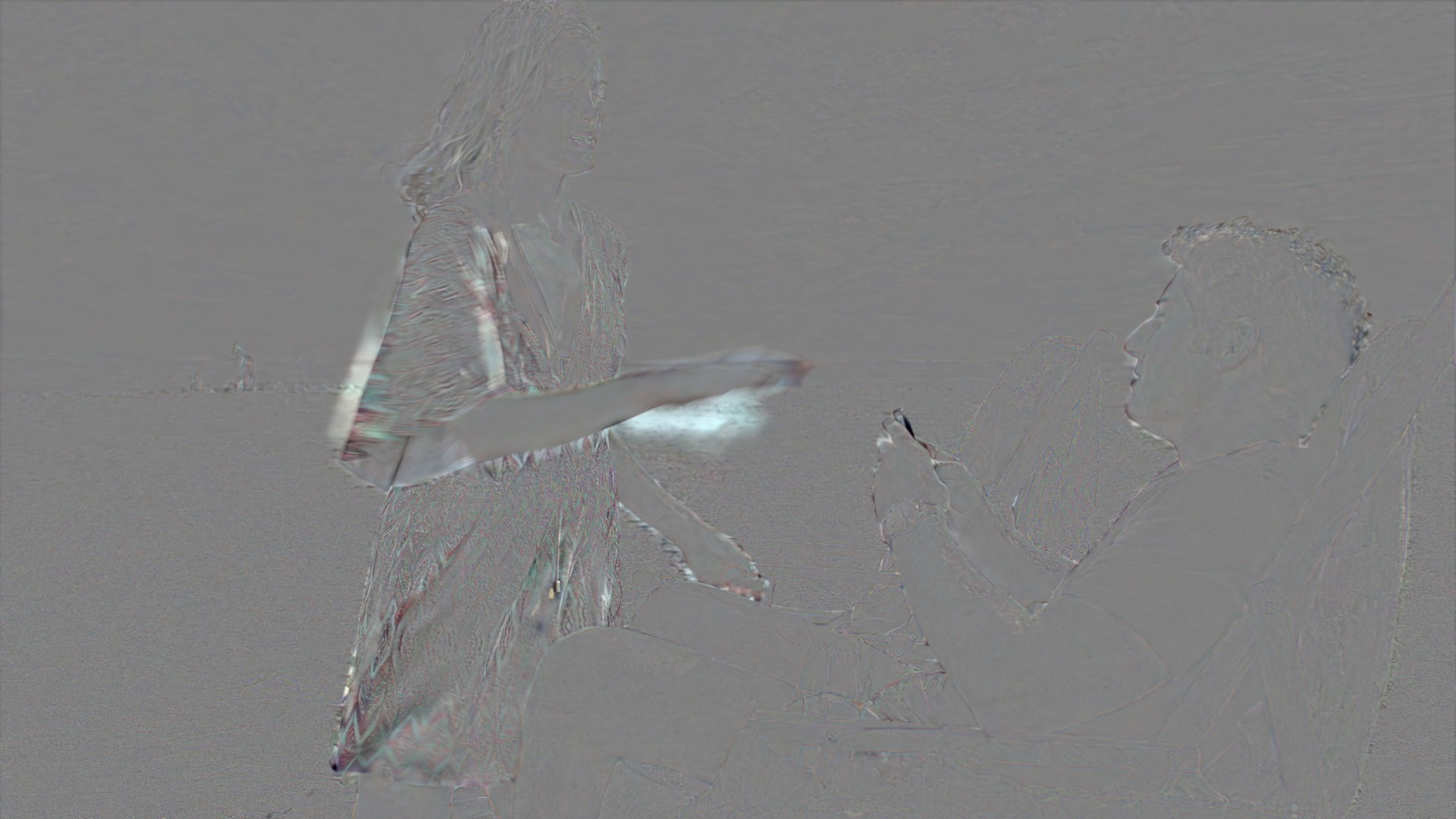}  \\[-0.75ex]
        $r_t$ 
        \end{tabular}
        }
    \caption{
    \label{fig:arch}
    Architecture overview, with some intermediate tensors visualized in the gray box. To the \emph{left} of the gray line is the I-frame branch (learned CNNs in blue), to the \emph{right} the P-frame branch (learned CNNs in green). Dashed lines are not active during decoding, and discriminators $\Dimg, \Dp$ are only active during training. The size of the CNN blocks roughly indicates their capacity. \texttt{SG} is a stop gradient operation. 
    \textit{DSSW} is our ``decouped scale-space warping'' (Sec.~\ref{sec:scalespaceblur}), and
    $\Uflow$ is a frozen optical flow model from~\cite{jonschkowski2020matters}.
    \vspace{-3ex}
    }
\end{figure}
\addtolength{\tabcolsep}{0.1ex}    

\mysection{Method}

\mysubsection{Overview}
\label{sec:architecture}

An overview of the architecture we use is given in Fig.~\ref{fig:arch}, while a detailed view with all layers is provided in App.~Fig.~\ref{fig:archdetail}. 
Let $x = \{x_1, x_2, \dots\}$ be a sequence of frames, where $x_1$ is the initial (I) frame, denoted by $x_I$ in the figure and below. Similar to previous work, we operate in the ``low-delay'' mode, and hence predict subsequent (P) frames from previous frames.
Let $\hat x = \{ \hat x_1, \hat x_2, \dots \}$ be the reconstructed video.
We use the following strategy to obtain high-fidelity reconstructions:
\setlist{nolistsep}
\begin{enumerate}[label=(S\arabic*),leftmargin=*,noitemsep]
    \item Synthesize plausible details in the I-frame. \label{strat:1}
    \item Propagate those details wherever possible and as sharp as possible. \label{strat:2}
    \item For new content appearing in P-frames, we again want to synthesize plausible details.  \label{strat:3}
\end{enumerate}
As mentioned in the Introduction, we optimize for perceptual quality and distortion, and note that the above three points are in contrast to purely distortion-optimized neural video codecs, which, particularly at low bitrates, favor blurring detail to reduce the distortion loss. 
Instead, our approach will be able to synthesize faithful texture, while \emph{still staying close to the input}, as seen in Fig.~\ref{fig:frontpic}.

The \textbf{I-frame branch} is based on a lightweight version  of the architecture used in ``HiFiC''~\cite{mentzer2020high} (mostly making it less wide, see App.~Fig.~\ref{fig:archdetail}), and is used to address \ref{strat:1}. In detail, the encoder CNN $\Eimg$ maps the input image $x_I$ to a quantized latent $\yimg$, which is entropy coded using a hyperprior~\cite{minnen2018joint} (not shown in Fig.~\ref{fig:arch}, but which is detailed in App.~Fig.~\ref{fig:archdetail}). 
From the decoded $\yimg$, we obtain a reconstruction $\hat x_I$ via the I-generator $\Gimg$.
We use an I-frame discriminator $\Dimg$ that---following~\cite{mentzer2020high}---is conditioned on the latent $z_I$ (we elaborate on conditioning in Sec.~\ref{sec:formulation}).

The \textbf{P-frame branch} has two parts, an auto-encoder $\Eflow$, $\Gflow$ for the flow, and an auto-encoder $\Eres, \Gres$ for the residual, following previous video work (\eg~\cite{lu2019dvc,agustsson2020scale}, \etc).
To partially address \ref{strat:2}, similar to previous work, we employ a powerful
optical flow predictor network on the encoder side, \textbf{\Uflow}~\cite{jonschkowski2020matters}.
The resulting (backward) flow $F_t=\Uflow(x_t, x_{t-1})$ is fed to the flow-encoder $\Eflow$, which outputs the quantized and entropy-coded flow-latent $\yflow$. From the flow-latent, the generator $\Gflow$ predicts both a reconstructed flow $\hat F_t$, as well as a mask $\sigma_t$.
The mask $\sigma_t$ has the same spatial dimensions as $F_t$, with each value in $[0, \sigma_\text{max}]$. 
Together, $(\hat F_t, \sigma_t)$ are used for our \textbf{decoupled scale-space warping}, a variant of \textit{scale-space warping}~\cite{agustsson2020scale}, described in Sec.~\ref{sec:scalespaceblur}.
Intuitively, for each pixel, the mask $\sigma_t$ predicts how ``correct'' the flow at that pixel is (see the gray box in Fig.~\ref{fig:arch}). We first warp the previous reconstruction $\hat x_{t-1}$ using $\hat F_t$, then we use $\sigma_t$ to decide how much to blur each pixel.
In practice, we observe $\sigma_t$ predicts where new content that is not well captured by warping appears. Since the flow is in general relatively easy to compress, we employ shallow networks for $\Eflow$ and $\Gflow$ based on networks used in image compression~\cite{minnen2018joint}.
We denote the resulting warped and potentially blurred previous reconstruction with $\hat x_t^w$.

Finally, we calculate the residual $r_t = x_t - \hat x_t^w$ and compress it with the \textbf{residual auto-encoder} $\Eres, \Gres$. To address the last point above, \ref{strat:3}, we again employ the light version of the HiFiC architecture for $\Eres, \Gres$. 
However, we introduce one important component. 
We observe that $\Gres$ is not able to synthesize high-frequency details from the sparse residual latent $\Eres(r_t)$ alone.
However, we found that additionally feeding a \textbf{``free'' latent} extracted from the warped previous reconstruction $y_t^\text{free} = \Eimg(\hat x_t^w)$ significantly increased the amount of synthesized detail, possibly due to the additional information and context provided by $\hat x_t^w$.
Note that this latent does not need to be encoded into the bitstream because the decoder already has $\hat{x}_t^w$ and can compute $y_t^{\text{free}}$ directly (hence it is ``free''), and thus also does not need to be quantized. Instead, we concatenate it to $\Eres(r_t)$ as a source of information, forming $\yres{=}\text{concat}(
y_t^\text{free}
,
\Eres(r_t)
)$.

To train the P-frame branch, we employ a seperate P-frame discriminator $\Dp$, with the same architecture as $\Dimg$, conditioned on the generator input $\yres$.

\begin{figure}[t]
\centering
\adjincludegraphics[trim={0 0 {0.5\width} {0.5\height}},clip,width=0.3\linewidth]{%
  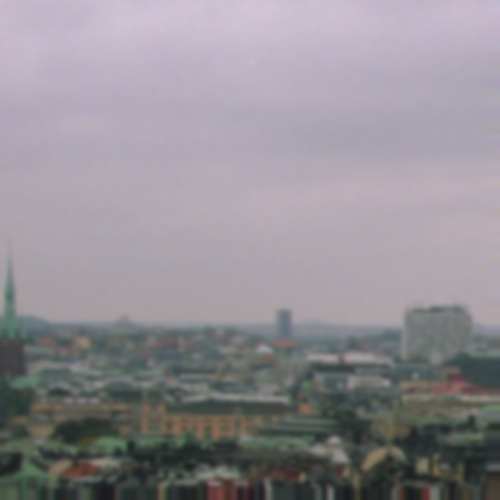}  
  \hspace{1em}
\adjincludegraphics[trim={0 0 {0.5\width} {0.5\height}},clip,width=0.3\linewidth]{%
  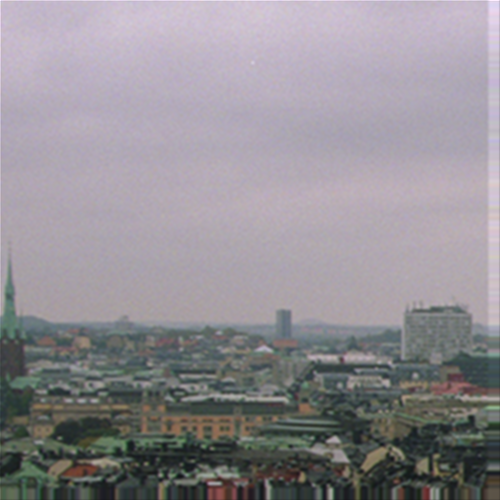} \\[-0.5ex]
{\footnotesize
\hfill Bilinear Resampling Kernel 
\hspace{1em}Bicubic Resampling Kernel \hfill}\vspace{-1ex}
\caption{\label{fig:bilinearbicubic}
To avoid blurry results when repeatedly warping, the quality of the resampling kernel is crucial. 
Here, we compare shifting an images 20 times with a fixed flow of 0.5px to the left for bilinear and bicubic.\vspace{-3ex}
}
\end{figure}

\addtolength{\tabcolsep}{-1ex}    
\begin{figure*}[t]
    \centering
    {\scriptsize
    \begin{tabular}{@{}r@{\hskip 0.8ex}l@{\hskip 0.1ex}l@{\hskip 0.1ex}l}
\multicolumn{1}{l}{\hphantom{j}Input $x_t$} &
\multicolumn{3}{l}{\hphantom{j}Output $\hat x_t$ for different models} \\
\includegraphics[trim={0 4em 0 5em},clip,width=0.246\textwidth]{%
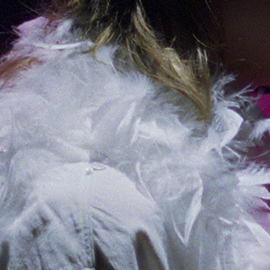} &
\includegraphics[trim={0 4em 0 5em},clip,width=0.246\textwidth]{%
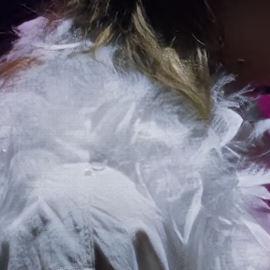} &
\includegraphics[trim={0 4em 0 5em},clip,width=0.246\textwidth]{%
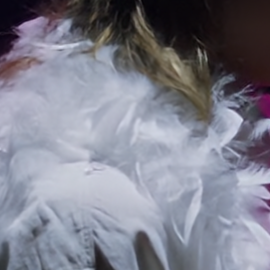} &
\includegraphics[trim={0 4em 0 5em},clip,width=0.246\textwidth]{%
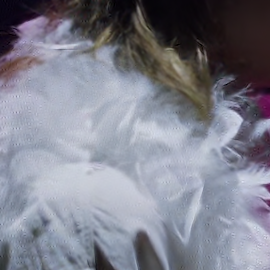} \\[-0.4ex]
& \name (with GAN & No GAN loss ($\beta{=}0$) & No free latent $y_t^\text{free}$ \\
&  and free latent)  \\[4ex]
\vspace{-0.7cm}%
        \end{tabular}}
    {\scriptsize
    \begin{tabular}{@{}r@{\hskip 0.8ex}l@{\hskip 0.1ex}l@{\hskip 0.1ex}l}
\multicolumn{1}{l}{\hphantom{j}Uncompressed Flow $F_t$} &
\multicolumn{3}{l}{\hphantom{j}Compressed Flow $\hat F_t$ for different models} \\
\includegraphics[trim={0 5em 0 8em},clip,width=0.246\textwidth]{%
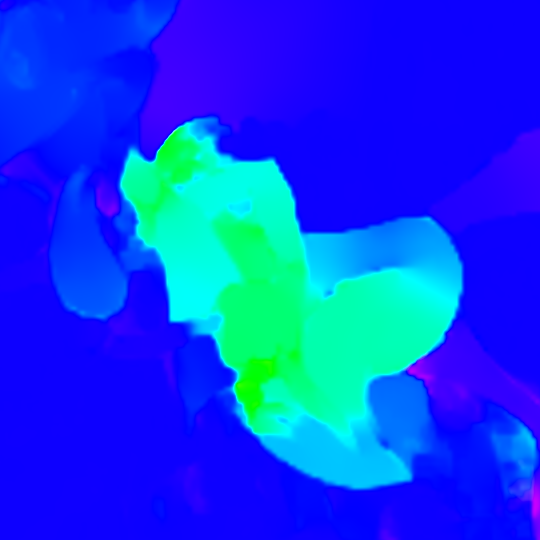} &
\includegraphics[trim={0 5em 0 8em},clip,width=0.246\textwidth]{%
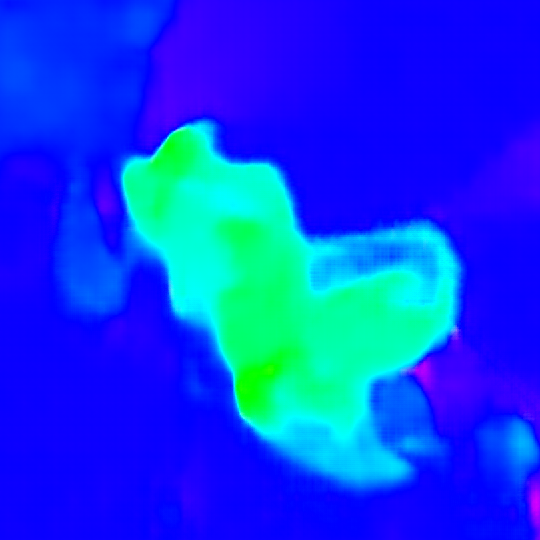} &
\includegraphics[trim={0 5em 0 8em},clip,width=0.246\textwidth]{%
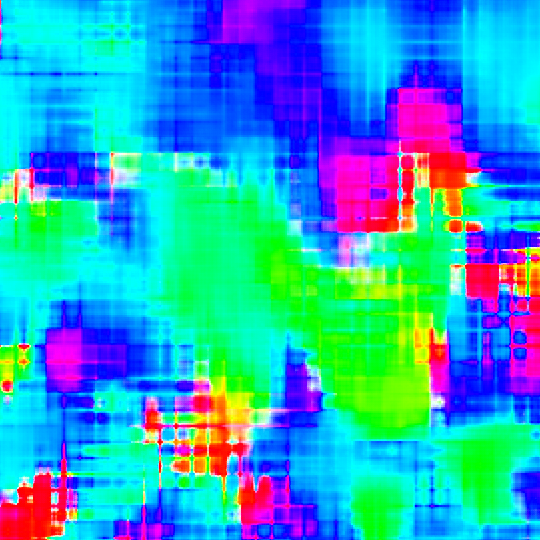} &
\includegraphics[trim={0 5em 0 8em},clip,width=0.246\textwidth]{%
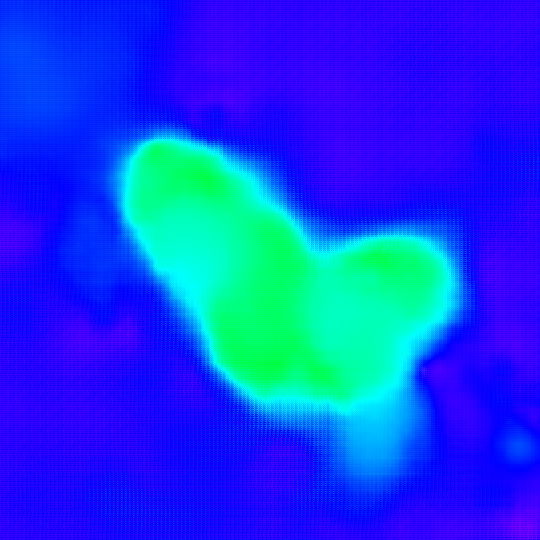} 
\\[-0.4ex]
& \name (with UFlow and & Without UFlow &
No flow loss ($k_\text{flow}{=}0$)\\
&  $k_\text{flow}{=}1$)
\end{tabular} \vspace{-2ex}
}
    \caption{\label{fig:uflow_cond_ablation}
    Visual examples for our ablations, see Sec.~\ref{sec:abl} for details.
    \emph{Top}: Our model faithfully reconstructs details of the input, whereas disabling the GAN loss or removing the free latent introduces blurryness like in MSE models. \emph{Bottom}: 
    Not using supervised optical flow (\Uflow) gives poor quality flows. Not using the flow loss makes the flow slightly burrier.\vspace{-3ex}
    }
\end{figure*}
\addtolength{\tabcolsep}{1ex}

\newcommand*{\Fhat}{\hat{F}}
\newcommand*{\xw}{x_{\text{warped}}}
\newcommand*{\xwb}{x_{\text{warped \& blurred}}}

\mysubsection{Decoupled Scale-Space Warping} \label{sec:scalespaceblur}
When warping previously reconstructed frames, we want to preserve detail as much as possible (whether real or synthesized, per \ref{strat:2} above). Previous neural video compression approaches have commonly used bi-linear warping \cite{lu2019dvc,djelouah2019neural,rippel2019learned,yang2020learning}, or tri-linear scale-space warping (SSW) \cite{agustsson2020scale,rippel2021elf,yang2020hierarchical}. However it is known from signal processing theory (see \eg Nehab~\etal~\cite[Fig.~10.6 on p.~64]{nehab2014fresh}) that for repeated  applications of re-sampling, the quality of the interpolation kernel is crucial to avoid low-pass filtering the signal and blurring out details. We visualize this phenomenon in Fig.~\ref{fig:bilinearbicubic}.

Motivated by these observations, we were interested in implementing the more powerful \textit{bicubic} warping in SSW, but found that this
makes the implementation significantly more complex when combined with the 3-D indexing of scale-space warping.
Instead, to be able to efficiently use bicubic warping (and arbitrary other warping operations), we propose a variant of scale-space warping~\cite{agustsson2020scale}, where we \textit{decouple} the operation into two steps: plain warping, followed by spatially adaptive blurring.
We can then use off-the-shelf warping implementations for the first part.

Both variants, at their core, use the scale-space \text{flow field}
$(\Fhat, \sigma)$, which generalizes optical flow $\Fhat$ by also specifying a ``scale'' $\sigma$, such that we get a triplet $(u_{ij}, v_{ij}, \sigma_{ij})$ for each target pixel $(i, j)$, where $u_{ij}, v_{ij}$ are the flow coordinates, and $\sigma_{ij}$ is the blurring scale to use.
We recall the method from~\cite{agustsson2020scale}:
To compute a \emph{scale-space warped} result
\begin{align}
    x_{\text{out}} = \text{SSW}(x, \Fhat, \sigma),
\end{align}
the source $x$ is first repeatedly convolved with Gaussian blur kernels to obtain a ``scale-space volume'' with $L$ levels,
\begin{equation}
V(x) = [x, x\ast G(s_1), \cdots, x \ast G(s_{L-1})],
\end{equation}
where $G(s_i)$ is the Gaussian blur kernel with std.\ deviation $s_i$, and $\{s_1, \dots, s_{L-1}\}$ are hyperparameters defining how blurry each level in the volume is.
The three coordinates of the scale-space flow field $(u_{ij}, v_{ij}, \sigma_{ij})$ are then used to \textit{jointly warp and blur} the source image, retrieving pixels via tri-linear interpolation from the scale-space volume.

We obtain a \textbf{Decoupled SSW (DSSW)} result by combining plain warping with spatially adaptive blurring (AB),
\begin{align}
   x_\text{out}' = \text{DSSW}(x, \hat F, \sigma) %
                 = \text{AB}(\text{Warp}(x, \hat F), \sigma),
\end{align}
where \emph{Warp} is plain warping, and \emph{AB} is \textit{functionally} the same as SSW with a zero flow, \ie
    $\text{AB}(y, \sigma) := \text{SSW}(y, 0, \sigma)$,
but can be implemented with a few lines of code using simple multiplicative masks for each level in the scale-space volume to apply the 1-D linear interpolation for each pixel (code in App.~\ref{app:sec:scalespaceblur}).

Together, bicubic warping and adaptive blurring help to propagate sharp detail when needed, while also facilitating smooth blurring when needed (\eg, for focus changes in the video).
See App.~Fig.~\ref{fig:vis_ss_blur} for a visualization of how a given input and sigma field $\sigma_t$ get blurred via scale-space blur.

We found that on a GPU, DSSW using an optimized warping implementation and our AB was $2{-}3{\times}$ faster than a naive SSW implementation.
In App.~\ref{app:sec:scalespaceblur}, we validate our implementation by training models for MSE, and showing that DSSW with bilinear warping obtains similar PSNR as SSW, and DSSW with bicubic warping yields a better model.

\begin{figure*}
    \centering
    \includegraphics[width=\linewidth]{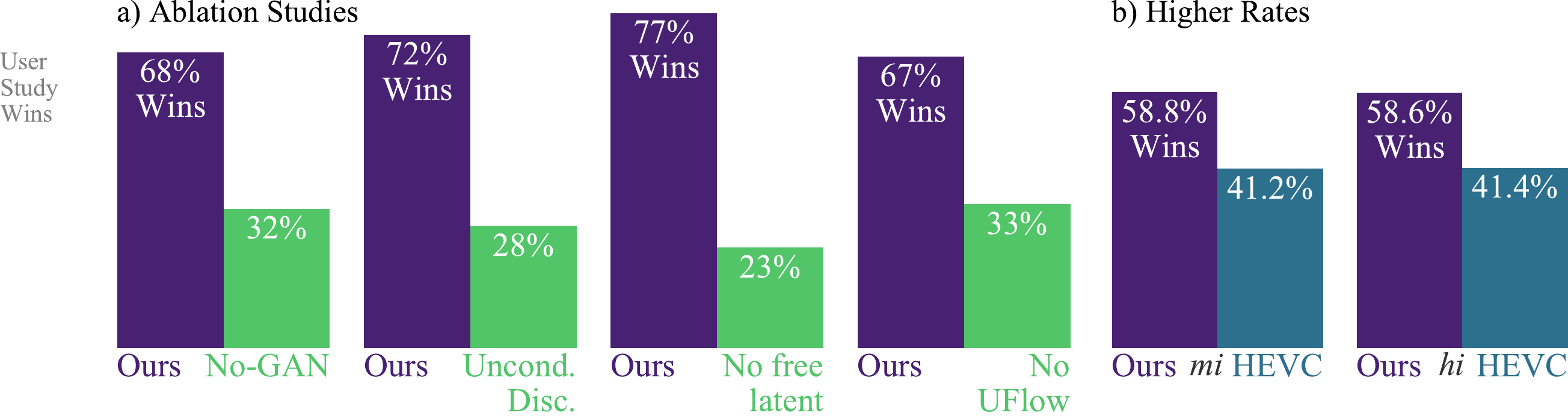}\vspace{-2ex}
    \caption{\label{fig:userstudyabl}\textit{a)} User study for Ablations. We see that disabling the GAN loss ($\beta{=}0$), using an unconditional discriminator, or not using the free latent hurts performance. On the flow side, not using UFlow hurts. 
    \textit{b)} Comparing models at higher rates, targeting 
    $0.14$bpp (\textit{mi}) and
    $0.22$bpp (\textit{hi}).\vspace{-3ex}
    }
\end{figure*}

\mysubsection{Adaptive Proportional Rate Control} \label{sec:rate-control}

\begin{figure}[b]
    \centering
    \includegraphics[width=0.8\linewidth]{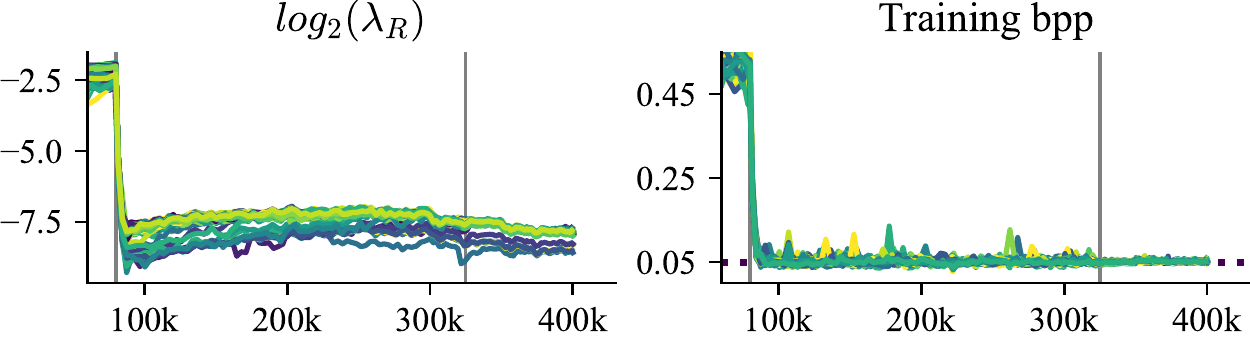}\vspace{-1ex}
    \caption{Visualizing the effect of the rate controller for a broad family of models with different hyper parameters, trained for 400k steps. The rate parameter $\lambda_R$ is automatically adapted during 
    training (\emph{left}) to match the target bpp of 0.05 for all models (\emph{right}). At 80k steps we drop the target rate, at 325k steps we drop the learning rate.
    \label{fig:rate_controller}
    }
\end{figure}

We train our system by optimizing the rate-distortion-perception trade-off~\cite{blau2019rethinking,mentzer2020high}, and we describe our formulation and loss in Sec.~\ref{sec:formulation}, but here we want to focus on
one hyper-parameter in this trade-off (that also typically appears in the rate-distortion trade-off optimized by previous work):
the weight on the bitrate, $\lambda_R$.
It controls the trade-off between bitrate and other loss terms (distortion, GAN loss, \etc). 
Unfortunately, since there is no direct relationship between $\lambda_R$ and the bitrate of the model when we vary other hyper-parameters, comparison across models is practically impossible, since they end up at different rates if we vary other hyper parameters, in particular other loss weights. 

Van Rozendaal~\etal~\cite{van2020lossy} also observe this and propose targeting a fixed distortion via constraint optimization. Another approach was used in~\cite{mentzer2020high}, where $\lambda_R$ was dynamically selected from a small $\lambda_1$ and a large $\lambda_2$, depending on whether the model bitrate was below or above a given target.
This approach can be interpreted as an ``on-off'' controller, but still some requires tuning of $\lambda_1, \lambda_2$.

\addtolength{\abovedisplayskip}{-1.2ex}    
\addtolength{\belowdisplayskip}{-1.2ex}    
\addtolength{\abovedisplayshortskip}{-1.2ex}    
\addtolength{\belowdisplayshortskip}{-1.2ex}    
A natural generalization is to use a \textbf{proportional-controller}:
We measure the error between the current mini-batch bitrate $b$ to a target 
bitrate $b_t$ (in log-space), and 
apply it with a proportional controller to update  $\lambda_R$ as follows:
\begin{equation}
    \log_2(\lambda_R) \leftarrow \log_2(\lambda_R) + k_P ( \log(b + \epsilon) - \log(b_t+\epsilon) ),
\end{equation}
where $\epsilon=1\sce{-9}$ for stability and the ``proportional gain'' $k_P$ is a hyperparameter.
We note that if we ignore the log-reparameterization, this corresponds to the ``Basic Differential Multiplier Method''~\cite{platt1988constrained}.
\addtolength{\abovedisplayskip}{1.2ex}    
\addtolength{\belowdisplayskip}{1.2ex}    
\addtolength{\abovedisplayshortskip}{1.2ex}    
\addtolength{\belowdisplayshortskip}{1.2ex}    

This approach is highly effective to obtain models that are comparable in terms of bitrate, despite different hyper-parameters such as learning rates, amount of unrolling, loss weights, \etc, as visualized in Fig.~\ref{fig:rate_controller}. 

\mysubsection{Sequence Length Train/Test Mismatch}\label{sec:unrolling}

One problem in neural video compression is the train/test mismatch in 
sequence length: Typically, neural approaches are trained on a handful of frames (\eg, three frames for~\cite{agustsson2020scale} and five for~\cite{liu2019neural}), and evaluated on hundreds of frames, which can lead to error patterns that emorge during evaluation.
While the unrolling behavior is already a problem for MSE-optimized neural codecs (some previous works use small GOPs of 8-12 frames for evaluation to limit temporal error propagation), it requires even more care when detail is synthesized in a generative setting. Since we aim to synthesize high-frequency detail whenever new content appears, incorrectly propagating that detail will create significant visual artifacts. 
Ideally, we could train with sequences as long as what we evaluate on (\ie, at least $T{=}60$ frames), but in
practice this is infeasible on current hardware due to memory and computational constraints.
While we can fit up to $T{=}12$ into our accelerators, training then becomes prohibitively slow. 

To work towards preventing unrolling issues,
as well as accelerating prototyping and training new models,
we instead adopt the following scheme: 1) Train $\Eimg, \Gimg, \Dimg$ only, on randomly selected frames, for $1\,000\,000$ steps. 2) Freeze $\Eimg, \Gimg, \Dimg$ and initialize the weights of $\Eres, \Gres$ from $\Eimg, \Gimg$. 
Train $\Eflow$, $\Gflow$, $\Eres$, $\Gres$, $\Dp$ for $400\,000$ additional steps using \textbf{staged unrolling}, that is, use $T{=}2$ until $80k$ steps, $T{=}3$ until $300k$, $T{=}4$ until $350k$, and $T{=}6$ until $400k$.
This splitting into steps 1) and 2) means trained $\Eimg, \Gimg$ can be re-used for many variants of the P-frame branch, and, as a bonus, sharing $\Eimg, \Gimg$ across models makes them more comparable. For training times, see App.~\ref{abl:sec:trainingtime}.

Some error accumulation remains, which we address in two ways: We quantize the frame buffer at each step, \ie, during inference, we always quantize $\hat x_t$, to be closer to the (8-bit quantized) input. Additionally, we randomly shift reconstructions in each step, to avoid overlapping larger-scale error patterns from accumulating. Together, these techniques help to get rid of most error patterns. %

\mysubsection{Formulation and Loss} \label{sec:formulation}

We base our formulation on HiFiC~\cite{mentzer2020high} and optimize the rate-distortion-perception trade-off~\cite{blau2019rethinking}.
We use conditional GANs~\cite{goodfellow2014generative,mirza2014conditional}, where both the generator and the discriminator have access to additional labels. As a short recap,
the general conditional GAN formulation assumes data points $x$ and labels $s$ following some joint distribution $p(x,s)$. 
The generator is supposed to map samples $s \sim p(s)$ to the distribution $p(x|s)$, 
and the discriminator is supposed to predict whether a given pair $(x,s)$ is from the real distribution $p(x|s)$, or from the generator distribution $p(\hat x|s)$. 

In contrast to HiFiC, we are working with \emph{sequences} of frames and reconstructions,
however, we aim for per-frame distribution matching, \ie, for $T$-length video sequences, the goal is to obtain a model s.t.:
\begin{align}
    p(\hat x_t | y_t) = p(x_t | y_t) \quad \forall t \in \{1, \dots, T\}, \label{eq:per-frame-dist-match}
\end{align}
where $x_t$ are inputs, $\hat x_t$ reconstructions (as above), and we condition both the generators and the discriminators on latents $y_t$, using $y_1{=}\yimg$ for the I-frame, $y_t{=}\yres$ for P-frames ($t{>}1$).
To readers more familiar with conditional generative video synthesis (\eg, Wang~\etal~\cite{wang2018vid2vid}), this simplification may seem sub-optimal as it may lead to temporal consistency issues (\ie, you may imagine that reconstructions $\hat x_t, \hat x_{t+1}$ are inconsistent).
We emphasize that since we are doing compression, we will also have a per-frame distortion loss (MSE), and we have information that we transmit to the decoder via a bitstream. So while the residual generator can in theory produce arbitrarily inconsistent reconstructions, in practice, these two points appear to be sufficient for preventing any temporal inconsistency issues in our models.  
We nevertheless explored variations where the discriminator is based on more frames, but this did not significantly alter reconstructions. 

Continuing from Eq.~\ref{eq:per-frame-dist-match}, we define the overall loss for the I-frame branch and its discriminator $\Dimg$ as follows. 
We use the ``non-saturating'' GAN loss~\cite{goodfellow2014generative}.
To simplify notation, let $\yimg = \Eimg(x_I), \hat x_I = \Gimg(\yimg)$:%
\begin{align}
\vsqueeze
    \loss{\textit{I-Frame}}&= 
      \mathbb{E}_{x_I \sim p(x_I)}\bigl[ 
        \lambda_R^I r(\yimg) + 
        d(x_I, \hat x_I) - 
        \beta \log(\Dimg(\hat x_I, \DcondI)) 
     \bigr] \label{eq:lossiframe}, \\
    \loss{\Dimg}&= 
        \mathbb{E}_{x_I \sim p(x_I)} \bigl[
            -\log(1 - \Dimg(\hat x_I, y_I))  
            -\log(\Dimg(x_I, y_I))  \bigr],
            \end{align}
where $\lambda_R^I$ is the adaptive rate controller described in Sec.~\ref{sec:rate-control},
$\beta$ is the GAN loss weight, and $d$ is a per-frame distortion.
We use $d{=}\text{MSE}$, \ie, in contrast to HiFiC~\cite{mentzer2020high}, we \textit{do not use a perceptual distortion} such as LPIPS. We found no benefit in training with LPIPS, possibly due to a more balanced hyper-parameter selection, and removing it speeds up training by ${\approx}35\%$.

For the P-frame branch, let $p(x_1^T)$ be the distribution of $T$-length clips, where we use $x_1$ as the I-frame, and let
\begin{align}
    \loss{\textit{P-Frame}} = 
    \mathbb{E}_{p(x_1^T)}&\bigl[  
    \textstyle \sum_{t=2}^T 
        \lambda_R^{P} r(\DcondP) + 
        t d(x_t, \hat x_t) 
        -t \beta \log(\Dp(\hat x_t, \DcondP)) +
        \loss{\text{reg}}
     \bigr] \label{eq:losspframe}, \\
    \loss{\Dp} = 
    \mathbb{E}_{p(x_1^T)}&\bigl[ 
    \textstyle \sum_{t=2}^T 
        -t \log(1 - \Dp(\hat x_t, \DcondP))
        -t \log(\Dp(x_I, \DcondP))  )
        \bigr] .
\end{align}

Note that we scale the losses of the $t$-th frame with $t$. This is motivated by the observation that $\hat x_t$ influences all $T-t$ reconstructions following it, and hence earlier frames indirectly have more influence on the overall loss. Scaling with $t$ ensures all frames have similar influence.

Additionally, we employ a simple regularizer for the P-frame branch:
\begin{align}
        \loss{\text{reg}} = 
            k_\text{flow} \cdot \texttt{SG}(\sigma_t) \cdot L_2(F_t, \hat F_t) + 
            k_\text{TV} \textit{TV}(\sigma_t), \label{eq:regularizer}
\end{align}
where the first part is an MSE on the flow, ensuring that $\Eflow, \Gflow$ learn to reproduce the flow from \Uflow. We mask it with the sigma field, since we only require consistent flow where the network actually uses the flow (but add a stop gradient, \texttt{SG}, to avoid minimizing the loss by just predicting $\sigma_t=0$).
\textit{TV} is a total-variation loss~\cite{shulman1989regularization} ensuring a smooth sigma field.

\mysection{Experiments}

\mysubsection{Datasets}
Our \textbf{training} data contains $992$k spatio-temporal crops of dimension $256{\times}256$, each containing 12 frames, obtained from videos from YouTube. For training, we randomly choose a contiguous sub-sequence of length $T\in\{2,3,4,5\}$, see Sec.~\ref{sec:unrolling}. The videos are filtered to originally be at least 1080p in resolution, 16:9, 30fps. We omit content labeled as ``video games'' or ``computer generated graphics'', using YouTube's category system~\cite{ytcat}.
We \textbf{evaluate} our model on the 30 videos of MCL-JCV~\cite{wang2016mcl}, which is available under a permissive license from USC, in contrast to, \eg, the HEVC test sequences, which are not publicly available. 
MCL-JCV contains a broad variety of content types and difficulty, including a wide variety of motion from natural videos, computer animation and classical animation. 

\mysubsection{User Study} \label{sec:userstudy}

\paragraph{2AFC} 
\begin{wrapfigure}[8]{r}{0.5\linewidth}
\centering
\vspace{-5ex}
\includegraphics[width=\linewidth]{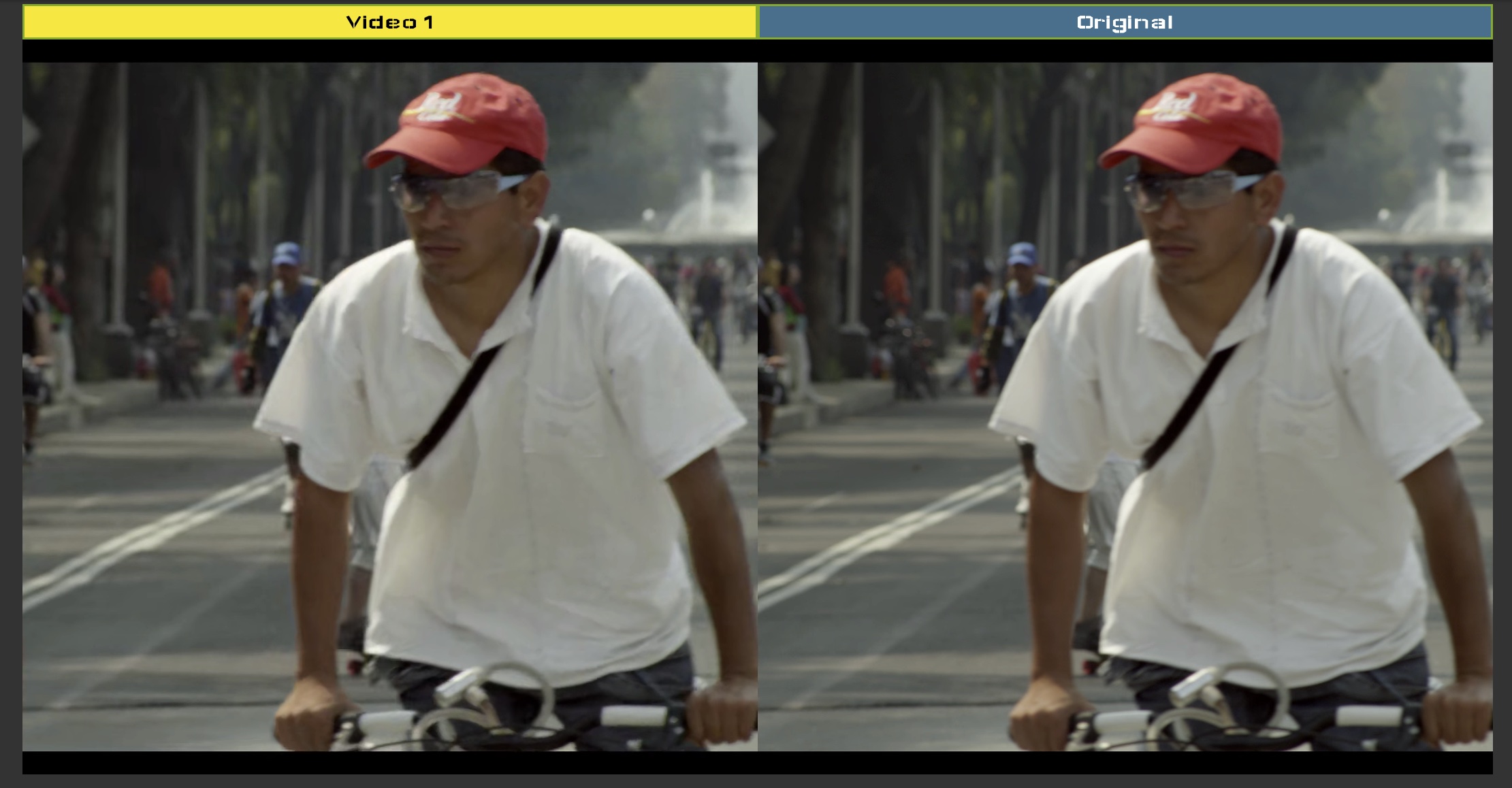}
\end{wrapfigure}
We evaluate our method in a user study, where we ask human raters to rate pairs of methods, \ie, our setup is ``two alternatives, forced choice'' (2AFC).
We implement 2AFC by showing raters two videos \textit{side-by-side}, where the right video is always the \textit{original}. On the left, raters see either a video from method A or method B. They can toggle between A and B in-place. We always shuffle the methods, \ie, \nameIT is not always method A.
We use all 30 videos from MCL-JCV, and show the first 2 seconds (to avoid large file sizes, see below), playing in a loop, but raters are allowed to pause videos.
Raters are asked to select the video ``is closest to the original''
(the GUI is shown in the inline figure, exact instructions in App.~Fig.~\ref{fig:rater_ui_instructions}). 
This protocol is inspired by previous work in image compression~\cite{mentzer2020high,clic2020}, and ensures that differences between methods are easy to spot.

Several considerations went into these choices:
For \textit{generative} video compression, it is important to  be able to compare to the original, as otherwise the method may, \eg, completely change colors or content. 
However, we do not require pixel-perfect reconstructions, which is why we show the original on the right, and not in-place.
Methods can be very similar, which is why we allow in-place switching \textit{between methods} to be able to spot differences.%

\paragraph{Rater Qualifications}
Our raters are contracted through the ``Google Cloud AI Labeling Service''~\cite{LabelingService}.
For each pair of methods, raters are asked to rate all 30 videos of MCL-JCV.
In order to make sure our ratings have a high quality, we intersperse \textbf{five golden questions} at random locations into each study, where we compare HEVC at quality factor $Q{=}27$ to $Q{=}35$ ($Q{=}27$ yields bitrates similar to what we study, and $Q{=}35$ is ${\approx}0.023$bpp and contains significant blurring artifacts). We filter out raters who do not correctly answer 4 out of 5 of these questions. Overall, this yields 8-14 raters per study. To ensure that our results are \textit{repeatable}, we re-do the study after three days.

\paragraph{Shipping Videos to Raters}
In order to play back the videos in a web browser we transcode all methods with VP9~\cite{mukherjee2013latest}, using a very high quality factor to avoid any new artifacts.
To ensure consistency, and to be sure that the raters can fit the task in their web browser, we center-crop the videos to $1080 \times 1080$.
This yields large file sizes, so to ensure smooth playback,
we \emph{focus on the first 2 seconds (60 frames) of each video}.%

\mysubsection{Metrics and Models} \label{sec:models}

In order to assess how quantitative metrics predict the user study, we employ the well-known PSNR and MSSSIM~\cite{wang2003multiscale}. We use LPIPS~\cite{zhang2018unreasonable} (which measures a distance in AlexNet~\cite{krizhevsky2012imagenet} features space) and PIM (an unsupervised quality metric), as well as FID~\cite{heusel2017gans}. 
Following HiFiC~\cite{mentzer2020high} we evaluate FID on non-overlapping $256{\times}256$ patches (see App.~A.7 in~\cite{mentzer2020high}).
Finally, we use VMAF~\cite{vmafurl}, developed by Netflix to evaluate video codecs.
We calculate these metrics on the exact sequences we ship to raters.

We refer to our model as \textbf{\name}, and report all hyper-parameters in App.~\ref{app:sec:hyperparameters}. 
To assess the effect of the GAN loss, we train a \textbf{\namebl} baseline, which uses exactly the same architecture and training schedule as \nameIT, but is trained without a GAN loss ($\beta{=}0$ in Eqs.~\eqref{eq:lossiframe},~\eqref{eq:losspframe}). We compare to three neural codecs: 
\textbf{\namecvpr}, by Agustsson~\etal~\cite{agustsson2020scale}, CVPR'2020, %
\textbf{RLVC} by Yang~\etal~\cite{yang2020learning}, J-STSP'2021,
and \textbf{DVC} by Lu~\etal~\cite{lu2019dvc}, CVPR'2019. 
For \namecvpr and RLVC, we obtained reconstructions from the authors, for DVC we ran the open-sourced code
ourselves and verified that this does match their published numbers on the UVG dataset~\cite{mercat2020uvg} (exact details in App.~\ref{app:sec:dvc}). 
We ran user studies comparing all these models against our proposed GAN model. 
In contrast to most previous work, we do not constrain our model to use a small GoP size, and instead only use an I-frame for the first frame.
For the neural models, we used the GoP from the respective papers ($\infty$ for SSF, 10 for DVC, 12 for RLVC), and we do not constrain the GoP for H.264 and HEVC.
The neural codecs we compare to do not densely cover the bitrate axis, so to ensure fair studies, we fix $\nameIT$ to a model targeting ${\approx}0.07$bpp, and then select a different competing model for each video to match filesizes as closely as possible. The resulting average bpps are at most ${\approx}3\%$ smaller or at most ${\approx}24\%$ bigger than our method.
We emphasize that we would have liked to compare to even more neural models, but found no additional code or reconstructions.

Furthermore,
we compare against the non-neural standard codecs \textbf{H.264} and \textbf{HEVC}.
We follow best practice and make sure to minimally constrain the codecs, thus using the default ``medium'' preset (note that some previous works used ``fast'' or even ``veryfast'').
Like our method, we run the codecs in the low-latency setting (disabling B-frames).
The exact (short) ffmpeg commands are listed in App.~\ref{app:sec:ffmpeg}.
We also run the codecs at ${\approx}0.07$bpp. To get an idea how models compare at higher rates, we fruther run \textit{HEVC} at ${\approx}0.14$ and ${\approx}0.22$bpp.%

\paragraph{On Padding and Bitrates}
A problem faced by all CNN-based neural compression codecs is: what happens if the stride of the network does not divide the input resolution. For example, our encoder downscales 4 times, and thus needs the input resolution to divide 16. Like most previous work, we solve this by \textit{padding} input frames (\eg, $1080{\times}1920$ gets padded to $1088{\times}1920$), obtaining the bitstream of the padded image, obtaining the reconstruction, cropping the reconstruction back to the input resolution, and \textit{calculating bpp w.r.t.\ the input resolution} (calculating it w.r.t.\ the padded resolution would amount to cheating). We note that the RLVC reconstructions were cropped to 1066 pixels, and we thus performed that user study in a cropped setting, and we had to add padding support to the DVC code, which may account for some differences in PSNR (DVC seems to have calculated on cropped images).

\begin{table}[t]
    \centering
{\small

\begin{tabular}{l@{\hskip 0.5ex}l@{\hskip 1ex}r@{\hskip 0.5ex}l@{\hskip 1.9ex}r@{\hskip 0.5ex}l@{\hskip 1.9ex}r@{\hskip 0.5ex}lr@{\hskip 0.5ex}l@{\hskip 1.9ex}r@{\hskip 0.5ex}l@{\hskip 1.9ex}r@{\hskip 0.5ex}l@{\hskip 0ex}}
\toprule
{} &  \rot{Ours} &  \rot{SSF} &                           \rot{Predicts?} &  \rot{RLVC} &                           \rot{Predicts?} &  \rot{DVC} &                     \rot{Predicts?} &  \rot{No-GAN} &                           \rot{Predicts?} &  \rot{H.264} &                           \rot{Predicts?} &  \rot{HEVC} &                           \rot{Predicts?} \\
\midrule
PSNR$\uparrow$      &        34.5 &       34.8 &  \textcolor{nored}{\textit{No$_\approx$}} &          34.0 &        \textcolor{yesgreen}{\textit{Yes}} &       31.7 &  \textcolor{yesgreen}{\textit{Yes}} &          35.1 &            \textcolor{nored}{\textit{No}} &         34.6 &  \textcolor{nored}{\textit{No$_\approx$}} &        35.6 &            \textcolor{nored}{\textit{No}} \\
MS-SSIM$\uparrow$   &       0.964 &      0.963 &  \textcolor{nored}{\textit{No$_\approx$}} &       0.965 &  \textcolor{nored}{\textit{No$_\approx$}} &       0.95 &  \textcolor{yesgreen}{\textit{Yes}} &         0.967 &  \textcolor{nored}{\textit{No$_\approx$}} &        0.963 &  \textcolor{nored}{\textit{No$_\approx$}} &       0.966 &  \textcolor{nored}{\textit{No$_\approx$}} \\
VMAF$\uparrow$      &        87.3 &       84.8 &        \textcolor{yesgreen}{\textit{Yes}} &        83.1 &        \textcolor{yesgreen}{\textit{Yes}} &       81.9 &  \textcolor{yesgreen}{\textit{Yes}} &          86.9 &  \textcolor{nored}{\textit{No$_\approx$}} &         87.7 &  \textcolor{nored}{\textit{No$_\approx$}} &        91.1 &            \textcolor{nored}{\textit{No}} \\
PIM-1$\downarrow$   &        3.34 &       4.69 &        \textcolor{yesgreen}{\textit{Yes}} &        4.93 &        \textcolor{yesgreen}{\textit{Yes}} &       6.91 &  \textcolor{yesgreen}{\textit{Yes}} &          4.17 &        \textcolor{yesgreen}{\textit{Yes}} &         3.17 &            \textcolor{nored}{\textit{No}} &        2.62 &            \textcolor{nored}{\textit{No}} \\
LPIPS$\downarrow$   &       0.168 &      0.224 &        \textcolor{yesgreen}{\textit{Yes}} &       0.224 &        \textcolor{yesgreen}{\textit{Yes}} &       0.26 &  \textcolor{yesgreen}{\textit{Yes}} &         0.194 &        \textcolor{yesgreen}{\textit{Yes}} &        0.169 &  \textcolor{nored}{\textit{No$_\approx$}} &       0.147 &            \textcolor{nored}{\textit{No}} \\
FID/256$\downarrow$ &        32.8 &       54.1 &        \textcolor{yesgreen}{\textit{Yes}} &        50.3 &        \textcolor{yesgreen}{\textit{Yes}} &       61.6 &  \textcolor{yesgreen}{\textit{Yes}} &          35.7 &        \textcolor{yesgreen}{\textit{Yes}} &           33.0 &  \textcolor{nored}{\textit{No$_\approx$}} &        24.2 &            \textcolor{nored}{\textit{No}} \\
\midrule 
\multicolumn{2}{l}{\scriptsize Preferred vs.\ Ours$\uparrow$}
 & \multicolumn{2}{c}{27\%} 
 & \multicolumn{2}{c}{31\%} 
 & \multicolumn{2}{c}{15\%} 
 & \multicolumn{2}{c}{32\%} 
 & \multicolumn{2}{c}{35\%} 
 & \multicolumn{2}{c}{39\%} 
 \\
\bottomrule
\end{tabular}

}

\caption{
    \label{tab:mainwithmetrics}
    We show metrics corresponding to the user studies, where the last row repeats the results from Fig.~\ref{fig:user_study_comparison}.
    We indicate whether each metric \emph{predicts} the study, using 
    \textcolor{yesgreen}{\textit{Yes}}
    and \textcolor{nored}{\textit{No}}.
    If the values are within 1\% of each other, the metric also \textit{does not predict} the study, and we indicate this with \textcolor{nored}{\textit{No$_\approx$}}.
    $\uparrow$ indicates that higher is better for this row, $\downarrow$ the opposite.
    We can see that no metric predicts all user studies (since \nameIT is preferred in all studies).
    \vspace{-4ex}
    }
\end{table}

\mysection{Results} \label{sec:results}

We show visual results in Fig.~\ref{fig:frontpic}. We can see how our approach faithfully synthesizes texture and looks very similar to the original, whereas MSE-based approaches suffer from blurryness. The quantitative results from our user study are shown in Fig.~\ref{fig:user_study_comparison}.
At a high level, we see that \nameIT is preferred by the majority in all studies. \name vs.\ \namebl shows that the GAN loss significantly improves visual quality. 
The first three studies show that our method significantly outperforms all neural baselines.
The standard codecs fare somewhat better, yet our method is clearly preferred overall. %
We show the comparison at higher rates in Fig.~\ref{fig:userstudyabl}b, where the gap between methods gets smaller, but our method is still preferred.

In Table~\ref{tab:mainwithmetrics} we explore which metrics are able to predict the user study results from Fig.~\ref{fig:user_study_comparison}. We show values of all methods on all metrics, and indicate whether the metrics predicts the corresponding study. 
\Eg,  we can see there that we are preferred over \nameblIT in the user study, yet our method has 34.5dB PSNR, while \nameblIT has 35.1dB (better), thus PSNR does \emph{not} predict this study correctly, and we write ``No''. 
Overall, none of the metrics are able to predict all studies. However, we find that the three ``perceptual'' metrics PIM, LPIPS, and FID/256 all predict the studies of the \textit{neural codecs}. Unfortunately, none of them predicts the studies involving the standard codecs. 

The table also shows how we trade-off distortion (PSNR) for improved realism/visual quality.
In the comparison against \nameblIT, we can see that 0.6dB in PSNR is traded for being preferred 68\% of the time in the user study.

In App.~\ref{app:sec:userstudy-details}, we show that we were able to obtain the same overall results when running the studies with the same raters three days later, with an even wider gap, and more raters passing the golden study. %
We also present statistics: how long raters take to answer questions, how often they flip, and how often they pause. We split this data by experiment, by video, and by worker. 
Averaged over all studies, raters take $26.4$s per comparison, flip $13.5$ times, and pause $0.967$ times.
To facilitate further research, we provide links to \textbf{reconstructions and raw user study data} in App.~\ref{app:sec:reconstruction-releases}.

\mysubsection{Ablations} \label{sec:abl}

We ablate our main components, using a user study (shown in Fig.~\ref{fig:userstudyabl}a) and visually (in Fig.~\ref{fig:uflow_cond_ablation}).
We do ablations by removing parts: 
In \textbf{No-GAN}, we disable the GAN loss ($\beta=0$),
for \textbf{No free latent} we train without the free latent $y_t^\text{free}$,
and 
in \textbf{Uncond.\ Disc.}, we train with an unconditional discriminator (\ie, $D$ does not see any latents).
We can see that all of these perform significantly worse in terms of visual quality (Fig.~\ref{fig:userstudyabl}), and lead to blurry reconstruction (Fig.~\ref{fig:uflow_cond_ablation}, uncond.\ disc.\ is not shown but looks similar to No-GAN).
In \textbf{No UFlow}, we disable $\Uflow$, \ie, if we do not feed $F_t$ to $\Eflow$, and instead let $\Eflow$ learn flow unsupervised from frames, which performs significantly worse (Fig.~\ref{fig:userstudyabl}). %

\mysection{Conclusion}

We presented a GAN-based approach to neural video compression, that significantly outperforms previous neural and non-neural methods, as measured in a user study.
With additional user studies, we showed that two components are crucial: i) conditioning the residual generator on a latent obtained from the warped previous reconstruction, and ii) leveraging accurate flow from an optical flow network. 
Furthermore, we showed how to decouple scale-space warping to be able to leverage high quality resampling kernels, and we used adaptive rate control to ensure consistent bitrates across a wide range of hyperparameters.
\paragraph{Limitations} As we saw, the quantitative metrics we currently have cannot be fully relied on, and hence we have to do user studies. However, this is expensive and not very scalable, and further research into perceptual metrics is needed. We hope that by releasing our reconstructions, we can encourage research in this direction.

\clearpage

{\footnotesize
\bibliographystyle{splncs04}
\bibliography{ref}
}

\clearpage
\appendix

\mysection{Appendix}

\mysubsection{HEVC/H.264 commands} \label{app:sec:ffmpeg}
We use \texttt{ffmpeg}, v4.3.2, to encode videos with HEVC and H264. We use the \texttt{medium} preset, and no B-frames, as mentioned in Sec~\ref{sec:models}. We compress each video using quality factors $\texttt{\$Q} \in \{10, \dots, 35\}$ to find bpps that match our models, using the following commands:

\noindent\begin{minipage}{\linewidth}
\vspace{1.2ex}
\begin{verbatim}
ffmpeg -i $INPUT_FILE -c:v h264 \
    -crf $Q -preset medium \
    -bf 0 $OUTPUT_FILE
\end{verbatim}

\noindent\end{minipage}
\begin{minipage}{\linewidth}
\begin{verbatim}
ffmpeg -i $INPUT_FILE -c:v hevc \
    -crf $Q -preset medium \
    -x265-params bframes=0 $OUTPUT_FILE
\end{verbatim}
\end{minipage}

\mysubsection{User Study Details}\label{app:sec:userstudy-details}

\begin{figure*}[hb]
    \centering
    
    \includegraphics[width=0.9\linewidth]{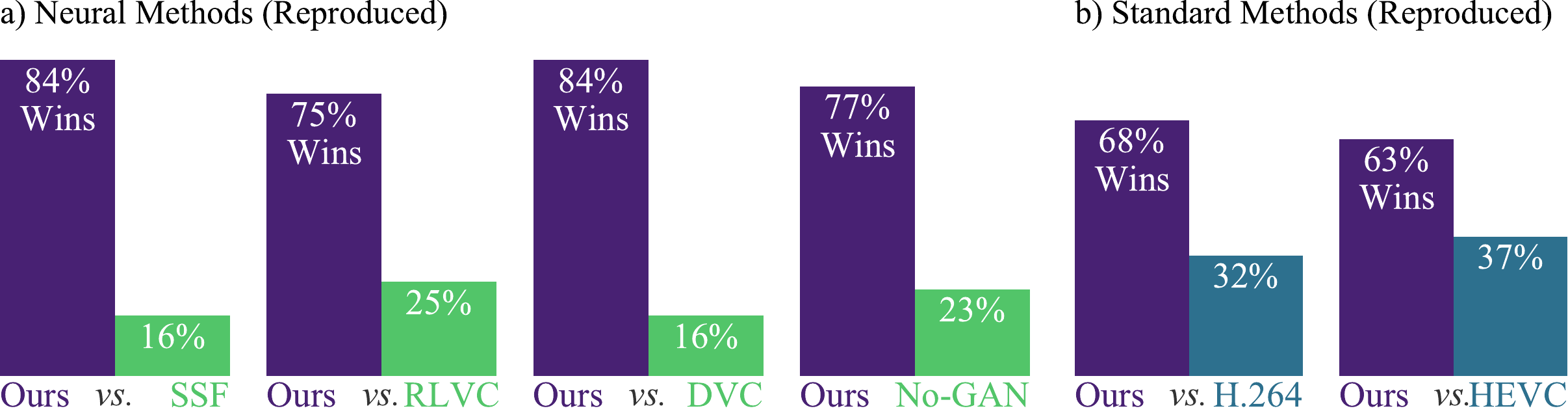}
    \includegraphics[width=0.85\linewidth]{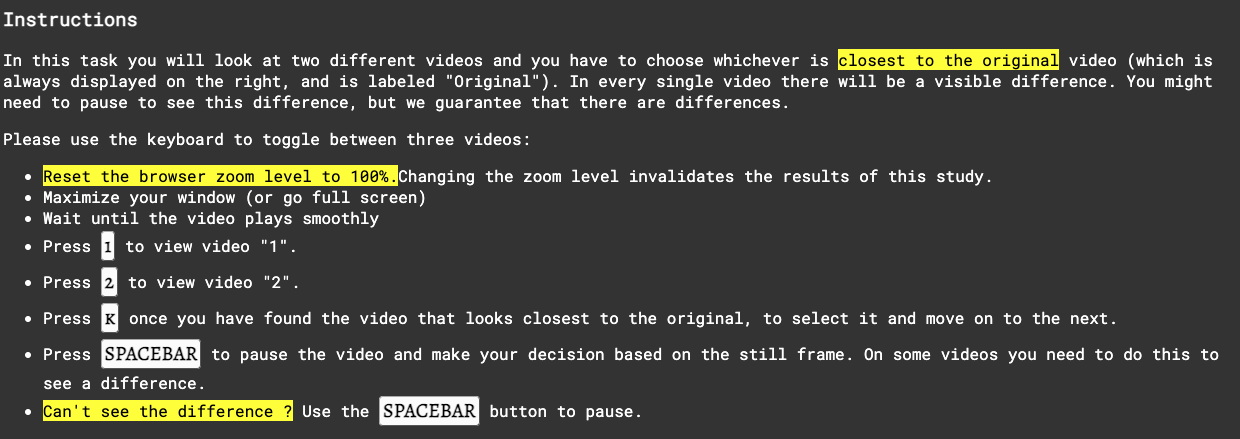}
    \caption{
    \textit{Top:} Study repeated 3 days later.
    \textit{Bottom:} Instructions shown to the raters.
    \label{fig:rater_ui_instructions}}
\end{figure*}

\begin{figure*}
    \centering
    \includegraphics[width=\linewidth]{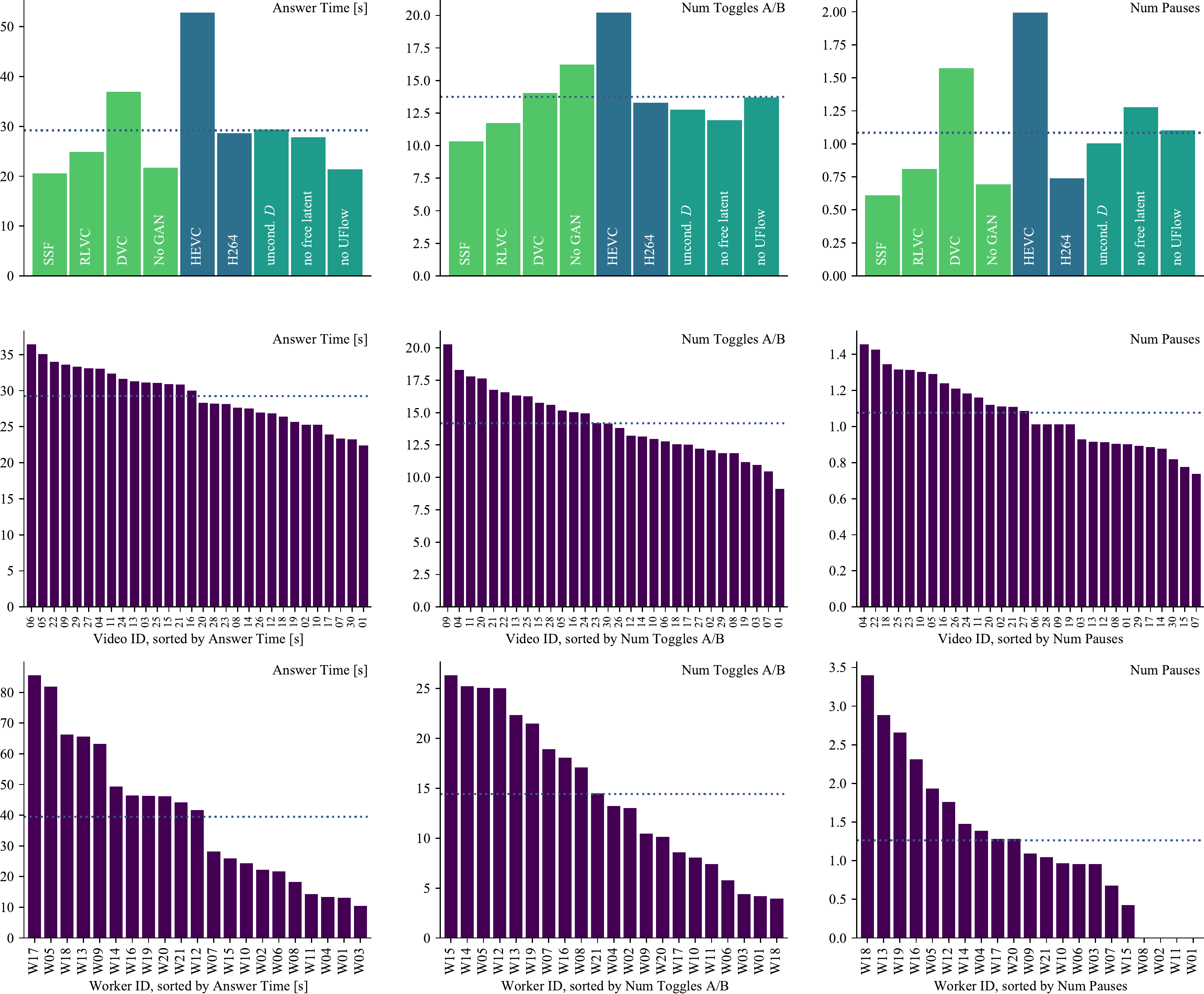}
    \caption{\label{app:fig:userstudyanalysis}
    User study statistics, grouped in various ways. 
    The dotted line in each plot indicates the mean of the values shown in the plot.
    From left to right, we show different statistics: Answer time in [s]econds, number of flips, and number of pauses. 
    The rows show different ways of grouping the data: \textit{Top}: We group by study, and color based on whether the study compares to a \collearned{\textit{neural codec}}, to a \colnonlearned{\textit{standard codec}}, or
    is an \colabl{\textit{ablation}}. 
    \textit{Middle}: We group by MCL-JCV video ID (01 to 30), and sort each plot.
    \textit{Bottom}: We group by rater ID, and sort each plot (note that the means here are slightly different, as not all raters rate did the same number of studies).
    }
\end{figure*}

\begin{itemize}[leftmargin=*]
\item Metrics for all ablation studies from Fig.~\ref{fig:userstudyabl} are shown in Table~\ref{tab:userstudyablmetrics}
\item The result of repeating the user studies three days later is shown in Fig.~\ref{fig:rater_ui_instructions} (top).
\item Rater instructions are shown in Fig.~\ref{fig:rater_ui_instructions} (bottom).
\item We show user study statistics in Fig.~\ref{app:fig:userstudyanalysis}, see caption for details.
\end{itemize}

\begin{table}[t]
    \centering
{\small

\begin{tabular}{l@{\hskip 0.5ex}r@{\hskip 1ex}l@{\hskip 0.5ex}l@{\hskip 1.9ex}r@{\hskip 0.5ex}l@{\hskip 1.9ex}r@{\hskip 0.5ex}l@{\hskip 1.9ex}r@{\hskip 0.5ex}l@{\hskip 1.9ex}r@{\hskip 0.5ex}l@{\hskip 1.9ex}r@{\hskip 0.5ex}l@{\hskip 0ex}}
\toprule
{} &  \rot{Ours} &  \rot{No-GAN} &                           \rot{Predicts?} &  \rot{Uncond. Disc.} &                           \rot{Predicts?} &  \rot{No free latent} &                           \rot{Predicts?} &  \rot{No UFlow} &                           \rot{Predicts?} &  \rot{HEVC} &                           \rot{Predicts?} &  \rot{HEVC} &                           \rot{Predicts?} \\
\midrule PSNR$\uparrow$      &        34.5 &          35.1 &            \textcolor{nored}{\textit{No}} &                 34.8 &  \textcolor{nored}{\textit{No$_\approx$}} &                  33.9 &        \textcolor{yesgreen}{\textit{Yes}} &            33.9 &        \textcolor{yesgreen}{\textit{Yes}} &          37 &            \textcolor{nored}{\textit{No}} &        38.2 &            \textcolor{nored}{\textit{No}} \\
MS-SSIM$\uparrow$   &       0.964 &         0.967 &  \textcolor{nored}{\textit{No$_\approx$}} &                0.966 &  \textcolor{nored}{\textit{No$_\approx$}} &                 0.959 &  \textcolor{nored}{\textit{No$_\approx$}} &            0.96 &  \textcolor{nored}{\textit{No$_\approx$}} &       0.974 &  \textcolor{nored}{\textit{No$_\approx$}} &       0.979 &  \textcolor{nored}{\textit{No$_\approx$}} \\
VMAF$\uparrow$      &        87.3 &          86.9 &  \textcolor{nored}{\textit{No$_\approx$}} &                 85.6 &        \textcolor{yesgreen}{\textit{Yes}} &                  81.9 &        \textcolor{yesgreen}{\textit{Yes}} &            84.2 &        \textcolor{yesgreen}{\textit{Yes}} &        94.7 &            \textcolor{nored}{\textit{No}} &        96.5 &            \textcolor{nored}{\textit{No}} \\
PIM-1$\downarrow$   &        3.34 &          4.17 &        \textcolor{yesgreen}{\textit{Yes}} &                 3.83 &        \textcolor{yesgreen}{\textit{Yes}} &                  3.85 &        \textcolor{yesgreen}{\textit{Yes}} &            3.32 &  \textcolor{nored}{\textit{No$_\approx$}} &        2.15 &            \textcolor{nored}{\textit{No}} &        1.75 &            \textcolor{nored}{\textit{No}} \\
LPIPS$\downarrow$   &       0.168 &         0.194 &        \textcolor{yesgreen}{\textit{Yes}} &                0.172 &        \textcolor{yesgreen}{\textit{Yes}} &                 0.194 &        \textcolor{yesgreen}{\textit{Yes}} &           0.167 &  \textcolor{nored}{\textit{No$_\approx$}} &       0.112 &            \textcolor{nored}{\textit{No}} &      0.0895 &            \textcolor{nored}{\textit{No}} \\
FID/256$\downarrow$ &        32.8 &          35.7 &        \textcolor{yesgreen}{\textit{Yes}} &                 34.9 &        \textcolor{yesgreen}{\textit{Yes}} &                  35.9 &        \textcolor{yesgreen}{\textit{Yes}} &            32.7 &  \textcolor{nored}{\textit{No$_\approx$}} &        15.5 &            \textcolor{nored}{\textit{No}} &        10.7 &            \textcolor{nored}{\textit{No}} \\
\midrule
\multicolumn{2}{l}{\scriptsize Preferred vs.\ Ours$\uparrow$}
&32\% &                                           &                 28\% &                                           &                  23\% &                                           &            33\% &                                           &        41.2\% &                                           &        41.4\%  \\
\bottomrule
\end{tabular}

}

\caption{
    \label{tab:userstudyablmetrics}
    We show metrics corresponding to the user studies, where the last row repeats the results from Fig.~\ref{fig:userstudyabl}.
    We indicate whether each metric \emph{predicts} the study, using 
    \textcolor{yesgreen}{\textit{Yes}}
    and \textcolor{nored}{\textit{No}}.
    If the values are within 1\% of each other, the metric also \textit{does not predict} the study, and we indicate this with \textcolor{nored}{\textit{No$_\approx$}}.
    $\uparrow$ indicates that higher is better for this row, $\downarrow$ the opposite.
    We can see that no metric predicts all user studies (since \nameIT is preferred in all studies).
    \vspace{-4ex}
    }
\end{table}

\clearpage

\mysubsection{Decoupled Scale-Space Warping: Details} \label{app:sec:scalespaceblur}

\begin{figure*}
    \centering
    \includegraphics[width=0.675\linewidth]{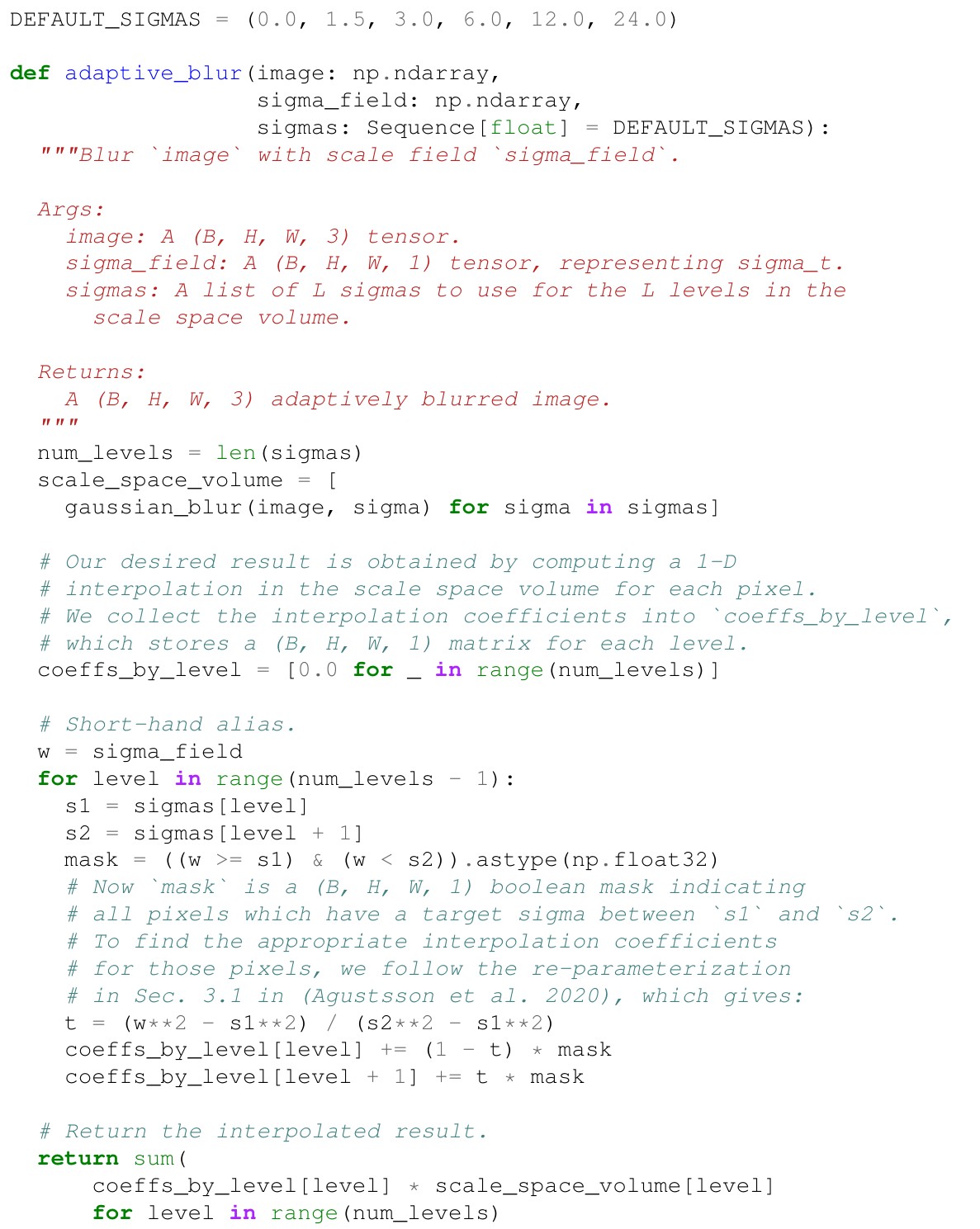}

    \caption{
    \label{fig:adaptive_blur_numpy}
    Numpy implementation of adaptive blurring.
    }
\end{figure*}

\begin{figure*}
\centering
{\small
\begin{tabular}{lll}
      \includegraphics[width=0.3\linewidth]{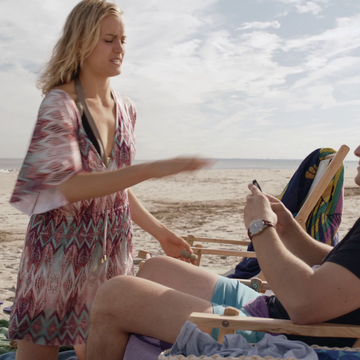} &
      \includegraphics[width=0.3\linewidth]{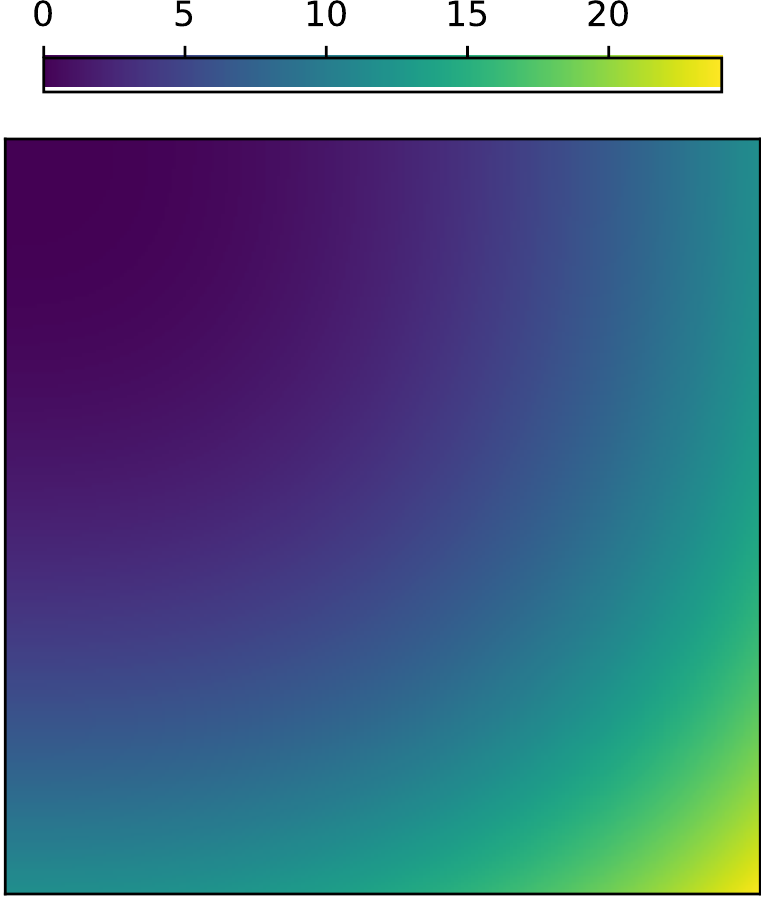} &
      \includegraphics[width=0.3\linewidth]{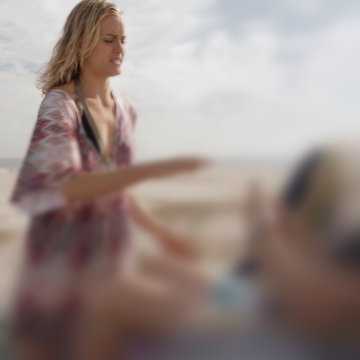} \\
      \texttt{image} &
      \texttt{sigma\_field} &
      \texttt{adaptive\_blur(image,} \\
      &&\hphantom{\texttt{xxxxxxxxxxxxxx}}\texttt{sigma\_field)} \\[2ex]
      \midrule  \\[2ex]
      \multicolumn{3}{l}{
          \texttt{coeffs\_by\_level[level]}} \\
      \includegraphics[width=0.3\linewidth]{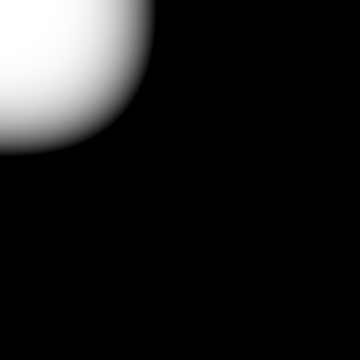} &
      \includegraphics[width=0.3\linewidth]{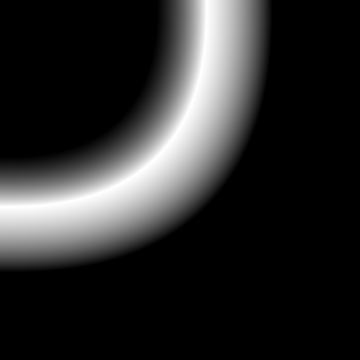} &
      \includegraphics[width=0.3\linewidth]{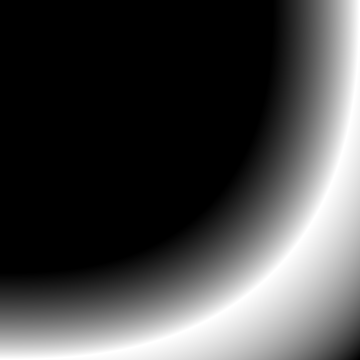} \\
      \texttt{level = 0} &
      \texttt{level = 2} &
      \texttt{level = 4} \\[2ex]
      \midrule  \\[2ex]
      \multicolumn{3}{l}{
          \texttt{coeffs\_by\_level[level] * scale\_space\_volume(image, sigmas[level])}} \\
      \includegraphics[width=0.3\linewidth]{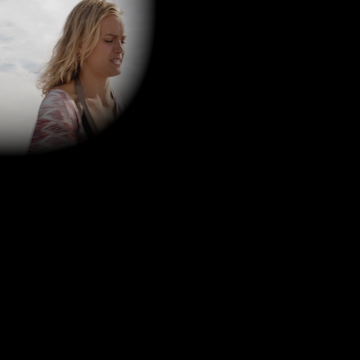} &
      \includegraphics[width=0.3\linewidth]{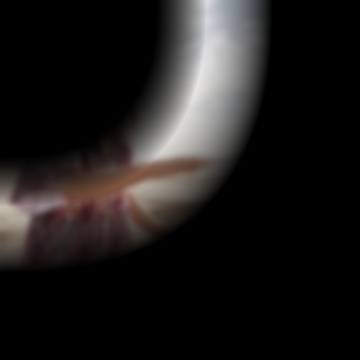} &
      \includegraphics[width=0.3\linewidth]{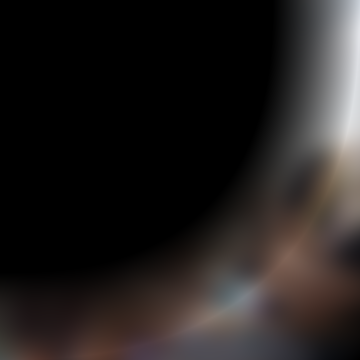} \\
      \texttt{level = 0} &
      \texttt{level = 2} &
      \texttt{level = 4} \\
\end{tabular}
}
\caption{\label{fig:vis_ss_blur}Visualizing variables of the algorithm given in Fig.~\ref{fig:adaptive_blur_numpy}.}
\end{figure*}

We show a numpy version of adaptive blurring in Fig.~\ref{fig:adaptive_blur_numpy},
and a visualization of some variables in Fig.~\ref{fig:vis_ss_blur}.
To validate our implementation of decoupled scale-space warping (DSSW), we compare MSE-trained models 
in Fig.~\ref{fig:ss_blur_vs_warp}.
We show that DSSW with bilinear warping is similar to scale-space warping with trilinear interpolation, validating our version. We see that using a bicubic resampling kernel, R-D performance improves by ${\approx}6\%$.
As we mention in Sec.~\ref{sec:scalespaceblur}, DSSW is also significantly faster to run on GPU.
For the Figure, we used the architecture of~\cite{agustsson2020scale} trained for MSE only, with a slightly accelerated training schedule by skipping the last training stage on larger (384px) crops, instead decaying the learning rate by 10x after 800\,000 steps.

\newpage

\noindent\begin{minipage}[b][6.5cm][b]{\linewidth}
\begin{center}
\includegraphics[width=0.5\linewidth]{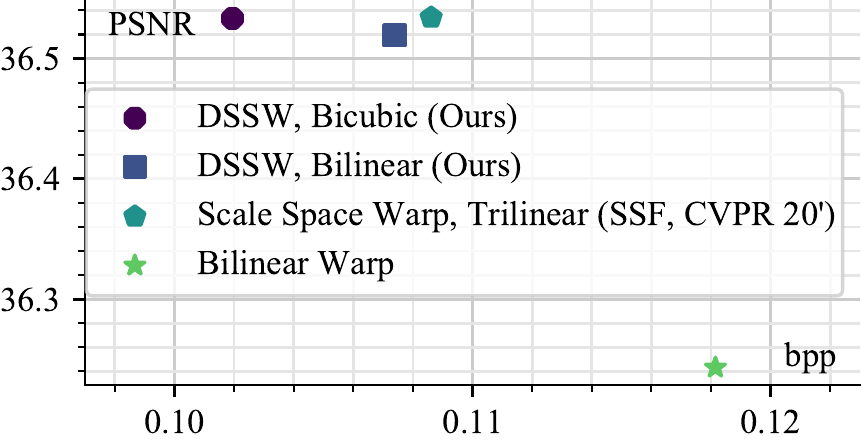}
    \captionof{figure}{%
    To validate our Decoupled Scale Space Warping (DSSW) implementation, we train models \textbf{for MSE}.
    We compare the R-D performance of Bilinear/Bicubic resamplers in DSSW against the Scale-Space Warping of Agustsson~\etal~\cite{agustsson2020scale} and find that DSSW with bicubic improves the bpp.
    We also show plain bilinear warping, without any scale space blurring.
    \label{fig:ss_blur_vs_warp}
    }
\end{center}
\end{minipage}

\newpage

\mysubsection{DVC Details} \label{app:sec:dvc}

\begin{figure}[t]
    \centering
    \includegraphics[width=0.5\linewidth]{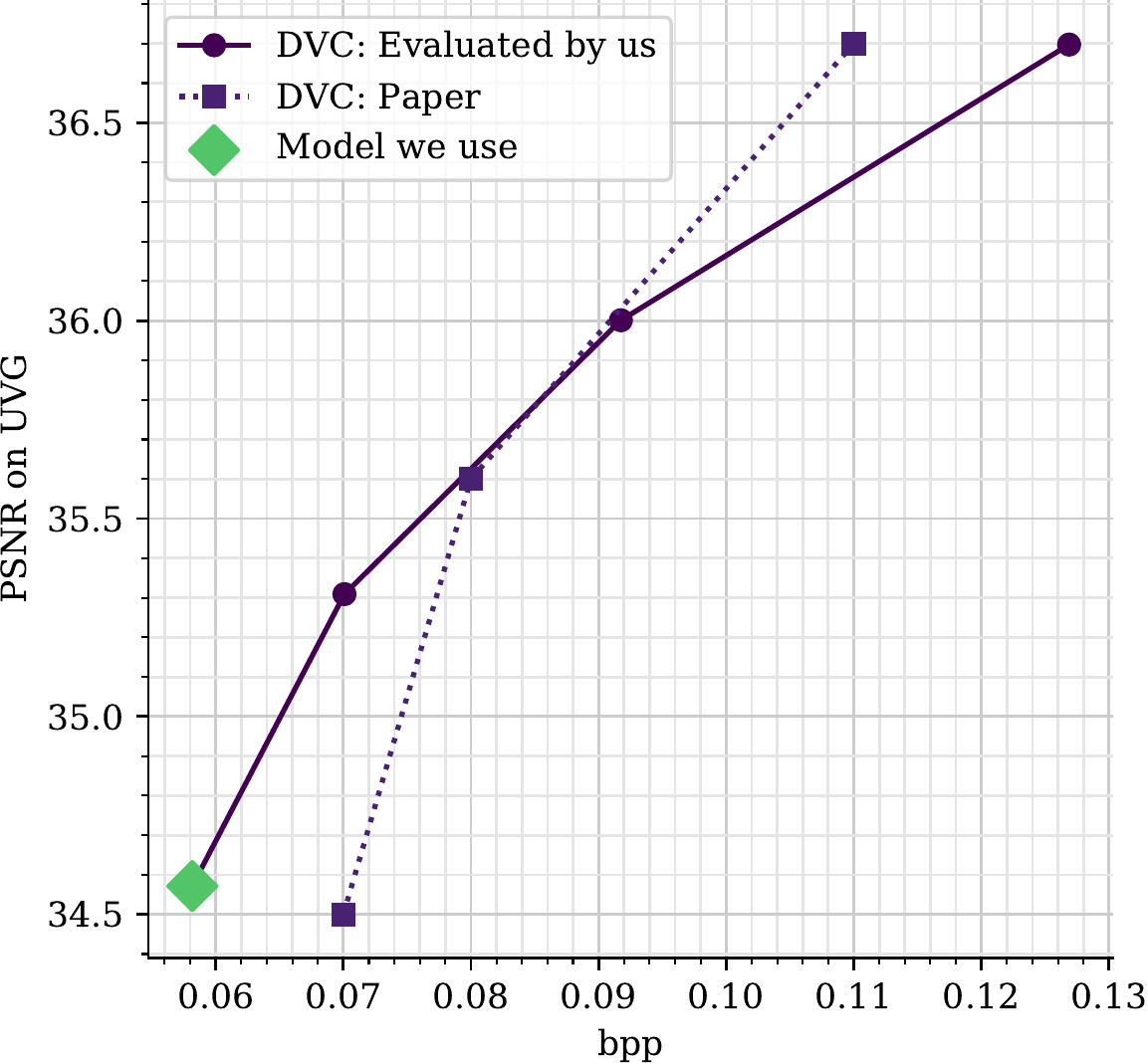}
    \caption{\label{fig:dvc}Comparing our DVC models to what the authors reported, on UVG. We use the model in the lower right, as this is closest to our bpps (achieving $\approx0.06$bpp on UVG, $\approx0.09$bpp on MCL-JCV).}
\end{figure}

To get DVC reconstructions, we use the code provided by the authors.\footnote{\url{https://github.com/GuoLusjtu/DVC}}
DVC uses the image compression model by Ballé~\etal~\cite{balle2018variational} for I-frames, but the code does not include the exact model. We thus tried all models, and picked the one with highest R-D performance, which is available as
``bmshj2018-hyperprior-mse-5'' in TFC.\footnote{\url{https://github.com/tensorflow/compression}}. We note that we add padding and cropping as described in Sec.~\ref{sec:models}. We show the PSNR of our model obtained on UVG in Fig.~\ref{fig:dvc}.

\begin{figure*}[t]
    \centering
    \includegraphics[width=1.0\textwidth]{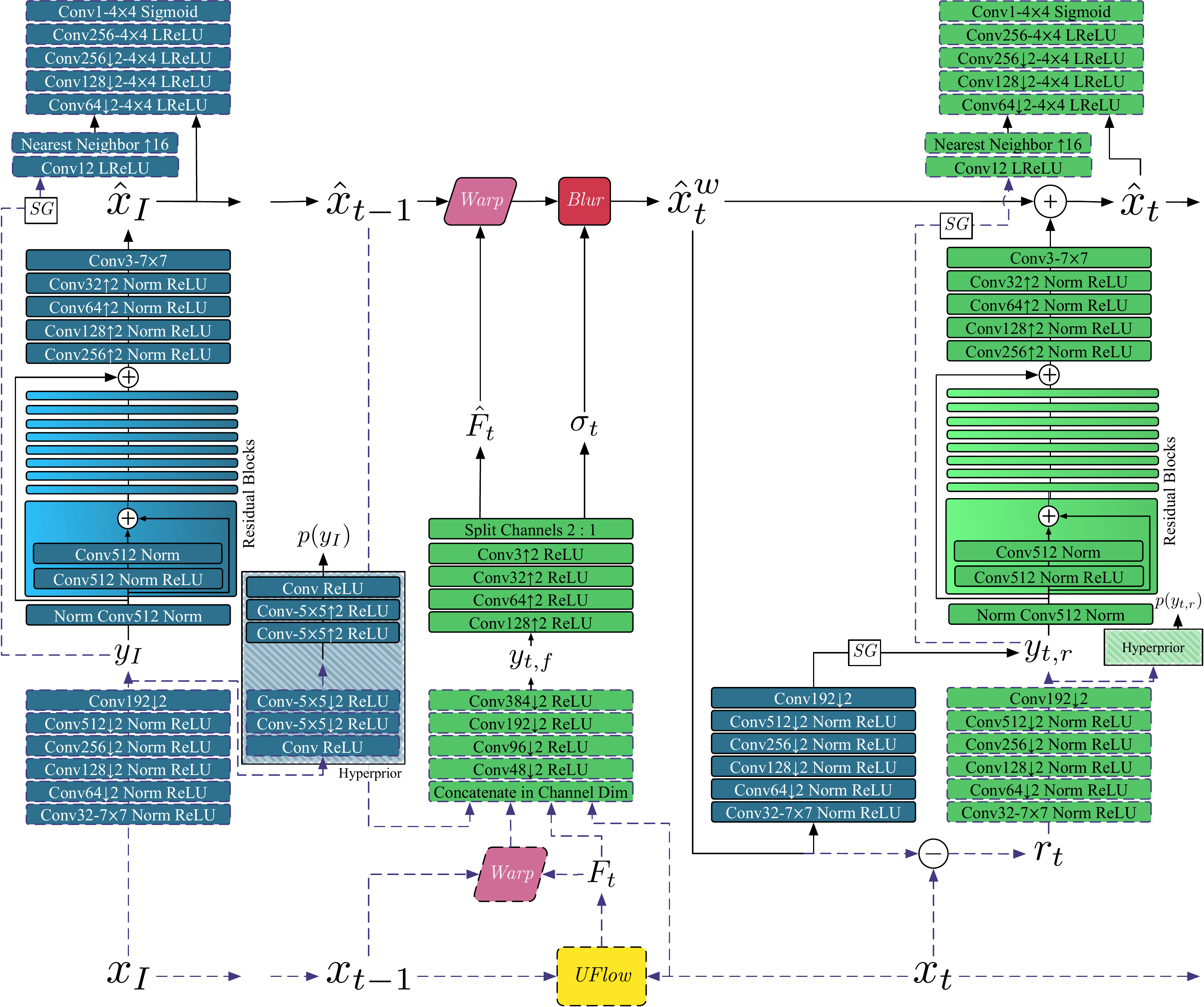}
    \caption{\label{fig:archdetail}
    Detailed view of the architecture, showing the layers in each of the blocks in Fig.~\ref{fig:arch}.
    ``Conv$F$'' denotes a 2D convolution with $F$ output channels, ``-S$\times$S'' denots the filter size, if that is omitted we use 3${\times}$3. ${\downarrow}2$, ${\uparrow}2$ indicates downsampling and upsampling, respectively, ``Norm'' is the \emph{ChannelNorm} layer employed by HiFiC~\cite{mentzer2020high}. The blocks with a color gradient are Residual Blocks, we only show the detail in one. ``LReLU'' is the Leaky ReLU with $\alpha=0.2$. We note that we employ SpectralNorm in both discriminators.
    The distributions predicted by the Hyperprior are used to encode the latents with entropy coding.
    Like in Fig.~\ref{fig:arch}, learned I-frame CNNs are in blue, learned P-frame CNNs in green, 
     dashed lines are not active during decoding, \texttt{SG} is a stop gradient operation, \textit{Blur} is scale space blurring, \textit{Warp} is bicubic warping. $\Uflow$ is a frozen optical flow model from~\cite{jonschkowski2020matters}.
     \vspace{-3ex}
     }
\end{figure*}

\mysubsection{Architecture Details} \label{app:sec:archdetails}
A detailed version of the architecture from Fig.~\ref{fig:arch} is given in Fig.~\ref{fig:archdetail}.

\mysubsection{Hyper Parameters}\label{app:sec:hyperparameters}

For scale space blurring we set $\sigma_0=1.5$ and used $L=6$ levels, which implies that the sequence of blur kernel sizes is $[0.0, 1.5, 3.0, 6.0, 12.0, 24.0]$.

For rate control we initially swept over a wide range $k_P \in 
\{10^{i}, i \in \{-1, \dots, -9\}\}$
and found that $10^{-3}$ worked well, which we then fixed for all future experiments. We initialized $\log_2 \lambda_R = 1.0$ in all cases.

Previous works~\cite{minnen2020channel,mentzer2020high} typically initially train for a higher bitrate. This is usually implemented by using a schedule on the R-D weight $\lambda$ that is decayed by a factor $2\times$ or $10\times$ early in training. 
Since the rate-controllor automatically controls this weight, we emulate the approach by instead using a schedule on the targeted bitrate $b_t$. We use a simple rule and target a $+0.5$ higher bitrate for the first $20\%$ of training steps.

For the I-frame loss $\loss{\textit{I-Frame}}$ (Eq.~\ref{eq:lossiframe}), we use $\beta = 128$, and $b_t=0.4$ for rate-control.

For the P-frame loss $\loss{\textit{P-Frame}}$ (Eq.~\ref{eq:losspframe}), we
use $\beta = 128$, and $k_\text{TV}=10.0, k_\text{flow}=1.0$ in $\loss{\text{reg}}$.
For the three different models we use in the user study, we use
$b_t \in \{0.05, 0.10, 0.15\}$.
A detail omitted from the equation is that we scale the loss by the constant $C_T = 1 / T \sum_{t=2}^T t$, as this yields similar magnitudes as no loss scaling.

We use the same learning rate $\text{LR}=1\sce{-4}$ for all networks, and train with the Adam optimizer. We linearly increase the LR from 0 during the first $20k$ steps, and then drop it to $\text{LR}=1\sce{-5}$ after $320k$ steps. We train the discriminators for 1 step for each generator training step. 

\mysubsection{Training Time} \label{abl:sec:trainingtime}
In Table~\ref{tab:training_speed} we report the training speed for each of the training stages, which results in a total training time of ${\approx}48$ hours. We note that the first stage (I-frame) trains more than $14{\times}$ faster than the last stage in terms of steps/s.

\begin{table}[b]
    \centering
\begin{tabular}{lllrrr}
\toprule
Batch size&   \#I &   \#P     &   \# steps [k]  &  steps/s & time [h] \\
\midrule
8        &      1        &      0            &       1\,000\,000       &  19.7     & 14.1  \\
8        &      1        &      1            &       80\,000         &  7.3      & 3.0 \\
8        &      1        &      2            &       220\,000        &  3.9      & 15.7 \\
8        &      1        &      3            &       50\,000         &  2.6      & 5.3 \\
8        &      1        &      5            &       50\,000         &  1.4      & 9.9 \\
\midrule
Totals:  &               &                   &     1\,400\,000         &           & 48.0 \\
\bottomrule
\end{tabular}
    \caption{Training speed/time for each stage of our model on a Google Cloud TPU. \#I, \#P indicates the number of I- resp.\ P-frames used in that stage.}
    \label{tab:training_speed}
\end{table}

\newpage

\clearpage

\mysection{Data Release} \label{app:sec:reconstruction-releases}

\mysubsection{CSVs and Reconstructions}

For each user study comparison we made between methods, we release the reconstructions as well as a CSV containing all the rater information, via anonymous links, see Table.~\ref{fig:appblinks}.
\begin{enumerate}[label=,leftmargin=0.1cm]
    \item \textit{Reconstructions folders:}
    \begin{enumerate}[label=]
        \item Folder per method, which contains a subfolder for each of the 30 videos of MCL-JCV, and each such video subfolder contains 60 PNGs, the reconstructions of the resp.\ method. 
    \end{enumerate}
    \item \textit{CSVs:}
    For each study, we release a CSV, where:
    \begin{enumerate}[label=]
        \item Each row is a video, and we have the following columns: \texttt{wins\_left}, \texttt{wins\_right} indicate the number of times each method won (left is always Ours),  \texttt{bpp\_left}, \texttt{bpp\_right}, indicate the per-video bpps, \texttt{avg\_flips}, \texttt{avg\_answer\_time\_ms}, \texttt{avg\_num\_pauses} indicate average flips, average time per video, and average num pauses, respectively.
    \end{enumerate}
\end{enumerate}

\begin{table*}
\begin{center}
\begin{varwidth}{0.8\textwidth}
\begin{enumerate}[align=right,leftmargin=1.5cm,itemsep=1ex]
\item[CSVs]{\url{https://storage.googleapis.com/eccv_sub/csvs.zip}}
\item[Ours]{\url{https://storage.googleapis.com/eccv_sub/ours.zip}}
\item[No-GAN]{\url{https://storage.googleapis.com/eccv_sub/nogan.zip}}
\item[SSF]{\url{https://storage.googleapis.com/eccv_sub/ssf.zip}}
\item[H.264]{\url{https://storage.googleapis.com/eccv_sub/h264.zip}}
\item[HEVC]{\url{https://storage.googleapis.com/eccv_sub/hevc.zip}}
\end{enumerate}
\end{varwidth}
\end{center}
    \caption{Links to user study data.}
    \label{fig:appblinks}
\end{table*}

\mysubsection{Tables}
We show wins per method per video, that are available in the CSVs, in Table~\ref{tab:winspermeth}.

\newpage

\begin{table}[h]
\centering

\begin{tabular}{rr@{\hskip 1ex}lr@{\hskip 1ex}lr@{\hskip 1ex}lr@{\hskip 1ex}lr@{\hskip 1ex}lr@{\hskip 1ex}l}
\toprule
\textbf{video} &
\rot{Ours} & \collearned{\rot{No-GAN}} & \rot{Ours} & \collearned{\rot{SSF}} & \rot{Ours} & \collearned{\rot{DVC}} & \rot{Ours} & \collearned{\rot{RLVC}} &
\rot{Ours} & \colnonlearned{\rot{H264}} & \rot{Ours} & \colnonlearned{\rot{HEVC}} \\
\midrule
\textbf{01   } &                        4 &                          4 &                        5 &                       4 &                       10 &                       1 &                        6 &                        3 &                        6 &                        2 &                        4 &                        6 \\
\textbf{02   } &                        4 &                          4 &                        7 &                       1 &                       10 &                       0 &                        4 &                        4 &                        2 &                        6 &                        3 &                        7 \\
\textbf{03   } &                        7 &                          2 &                        7 &                       1 &                       11 &                       0 &                        6 &                        2 &                        7 &                        2 &                        7 &                        3 \\
\textbf{04   } &                        5 &                          3 &                        5 &                       3 &                       11 &                       0 &                        5 &                        3 &                        8 &                        0 &                        8 &                        2 \\
\textbf{05   } &                        6 &                          2 &                        7 &                       1 &                        9 &                       1 &                        6 &                        3 &                        7 &                        2 &                        6 &                        4 \\
\textbf{06   } &                        3 &                          5 &                        5 &                       4 &                        7 &                       3 &                        6 &                        3 &                        4 &                        4 &                        8 &                        2 \\
\textbf{07   } &                        7 &                          1 &                        7 &                       1 &                       12 &                       0 &                        6 &                        2 &                        7 &                        1 &                        8 &                        2 \\
\textbf{08   } &                        7 &                          1 &                        7 &                       2 &                       10 &                       1 &                        6 &                        3 &                        5 &                        4 &                        9 &                        1 \\
\textbf{09   } &                        4 &                          4 &                        8 &                       2 &                        6 &                       5 &                        6 &                        3 &                        6 &                        2 &                        7 &                        3 \\
\textbf{10   } &                        5 &                          3 &                        8 &                       1 &                        8 &                       3 &                        7 &                        2 &                        5 &                        3 &                        2 &                        8 \\
\textbf{11   } &                        7 &                          1 &                        5 &                       3 &                       10 &                       1 &                        7 &                        2 &                        5 &                        3 &                        6 &                        4 \\
\textbf{12   } &                        7 &                          1 &                        7 &                       1 &                       11 &                       0 &                        5 &                        3 &                        4 &                        4 &                        7 &                        3 \\
\textbf{13   } &                        5 &                          3 &                        3 &                       5 &                        9 &                       2 &                        4 &                        5 &                        4 &                        5 &                        8 &                        2 \\
\textbf{14   } &                        6 &                          2 &                        7 &                       2 &                        9 &                       2 &                        7 &                        2 &                        6 &                        2 &                        8 &                        2 \\
\textbf{15   } &                        7 &                          2 &                        8 &                       1 &                        8 &                       2 &                        7 &                        3 &                        7 &                        1 &                        8 &                        2 \\
\textbf{16   } &                        4 &                          4 &                        7 &                       2 &                        9 &                       2 &                        5 &                        4 &                        6 &                        2 &                        8 &                        2 \\
\textbf{17   } &                        4 &                          4 &                        5 &                       3 &                       10 &                       2 &                        6 &                        2 &                        7 &                        1 &                        9 &                        1 \\
\textbf{18   } &                        5 &                          3 &                        6 &                       2 &                       10 &                       1 &                        7 &                        2 &                        6 &                        3 &                        6 &                        4 \\
\textbf{19   } &                        7 &                          2 &                        8 &                       1 &                        8 &                       2 &                        8 &                        1 &                        5 &                        3 &                        6 &                        4 \\
\textbf{20   } &                        6 &                          2 &                        3 &                       6 &                       10 &                       0 &                        8 &                        0 &                        1 &                        8 &                        3 &                        7 \\
\textbf{21   } &                        3 &                          5 &                        4 &                       5 &                        8 &                       3 &                        6 &                        3 &                        4 &                        4 &                        3 &                        7 \\
\textbf{22   } &                        5 &                          3 &                        5 &                       3 &                        9 &                       3 &                        5 &                        4 &                        6 &                        2 &                        7 &                        3 \\
\textbf{23   } &                        5 &                          3 &                        7 &                       2 &                       10 &                       1 &                        8 &                        2 &                        6 &                        2 &                        8 &                        2 \\
\textbf{24   } &                        5 &                          3 &                        6 &                       2 &                       10 &                       1 &                        6 &                        4 &                        5 &                        3 &                        2 &                        8 \\
\textbf{25   } &                        6 &                          2 &                        8 &                       1 &                        8 &                       3 &                        5 &                        5 &                        5 &                        3 &                        5 &                        5 \\
\textbf{26   } &                        4 &                          4 &                        6 &                       2 &                        8 &                       2 &                        8 &                        1 &                        4 &                        5 &                        7 &                        3 \\
\textbf{27   } &                        8 &                          0 &                        7 &                       1 &                       10 &                       1 &                        6 &                        3 &                        7 &                        1 &                        8 &                        2 \\
\textbf{28   } &                        7 &                          1 &                        7 &                       1 &                        9 &                       2 &                        6 &                        3 &                        8 &                        0 &                        5 &                        5 \\
\textbf{29   } &                        7 &                          1 &                        5 &                       4 &                        7 &                       4 &                        5 &                        4 &                        2 &                        7 &                        2 &                        8 \\
\textbf{30   } &                        5 &                          3 &                        8 &                       1 &                        9 &                       1 &                        7 &                        2 &                        6 &                        2 &                        6 &                        4 \\
\bottomrule\\[-3ex]
&\rotslant{165}&\rotslant{78}&\rotslant{188}&\rotslant{68}&\rotslant{276}&\rotslant{49}&\rotslant{184}&\rotslant{83}&\rotslant{161}&\rotslant{87}&\rotslant{184}&\rotslant{116}
\end{tabular}

\caption{\label{tab:winspermeth}Wins per method for our user studies.}
\end{table}

\end{document}